\documentclass{article}[11pt] 

\usepackage{amsmath} 
\usepackage{amssymb}
\usepackage{mathabx}
\usepackage[scr = dutchcal]{mathalpha}
\usepackage{tikz}
\usetikzlibrary{arrows,shapes,shapes.misc,shadows,shadings,decorations.markings,automata}

\tikzset{cross/.style={cross out, draw=blue, minimum size=2*(#1-\pgflinewidth), inner sep=0pt, outer sep=0pt},
cross/.default={1.5pt}}

\usepackage{verbatim}
\usepackage{changepage}
\usepackage{pgfplots}

\usepackage[hmarginratio=1:1,top=32mm]{geometry} 

\usepackage{enumitem} 

\usepackage{titling} 
\usepackage{amsmath}
\usepackage{amsthm}
\usepackage{hyperref} 

\usepackage{caption}
\usepackage[labelformat=simple]{subcaption}
\usepackage{graphicx}
\usepackage{xcolor}
\definecolor{red}{RGB}{146,0,0}
\definecolor{blue}{RGB}{0,109,219}
\definecolor{green}{RGB}{36,255,36}

\DeclareMathOperator{\Opt}{Opt}

\DeclareMathOperator{\NP}{NP}
\DeclareMathOperator{\ar}{ar}

\DeclareMathOperator{\Val}{Val}

\DeclareMathOperator{\tr}{tr}
\DeclareMathOperator{\Term}{\bf Term}
\DeclareMathOperator{\End}{End}

\DeclareMathOperator{\B}{B}

\DeclareMathOperator{\Irrep}{Irrep}

\DeclareMathOperator{\GL}{GL}
\DeclareMathOperator*{\E}{E}

\pretitle{\begin{center}\Huge\bfseries} 
\posttitle{\end{center}} 
\title{Max-3-Lin over Non-Abelian Groups with Universal Factor Graphs} 
\author{%
  {\textsc{Amey Bhangale}  } \\[1ex]  
\normalsize Department of Computer Science and Engineering
\\ \normalsize University of California, Riverside\\ 
\normalsize \href{mailto:ameyrb@ucr.edu}{ameyb@ucr.edu} 
\and 
{\textsc{Aleksa Stankovi\'c} \thanks{Research supported by the Approximability and Proof Complexity project funded by the Knut and Alice Wallenberg Foundation.}} \\[1ex] 
\normalsize Department of Mathematics \\ 
\normalsize KTH Royal Institute of Technology\\ 
\normalsize \href{mailto:aleksas@kth.se}{aleksas@kth.se} 
}
\date{\today} 
\renewcommand{\maketitlehookd}{%
\begin{abstract}
	Factor graph of an instance of a constraint satisfaction problem with $n$ variables and $m$ constraints is the bipartite graph between $[m]$ and $[n]$ describing which variable appears in which constraints. Thus, an instance of a CSP is completely defined by its factor graph and the list of predicates. We show inapproximability of  Max-$3$-LIN over non-abelian groups (both in the perfect completeness case and in  the imperfect completeness case), with the same inapproximability factor as in the general case, even when the factor graph is fixed.  
	
	Along the way, we also show that these optimal hardness results hold even when we restrict the linear equations in the Max-$3$-LIN instances to the form $x\cdot y\cdot z = g$, where $x,y,z$ are the variables and $g$ is a group element. We use representation theory and Fourier analysis over non-abelian groups to analyze the reductions.
\noindent 
\end{abstract}
}

\hypersetup{
  colorlinks,
  citecolor=black,
  linkcolor=red,
  urlcolor=blue}

\theoremstyle{plain}
\newtheorem{thm}{Theorem}[section] 
\newtheorem{lemma}[thm]{Lemma} 
\newtheorem{theorem}[thm]{Theorem} 

\theoremstyle{definition}
\newtheorem{definition}[thm]{Definition} 

\DeclareMathOperator{\OPT}{OPT}
\newcommand{\eps}{\varepsilon}

\begin{document}

\maketitle

\section{Introduction}

Constraint Satisfaction Problems (CSPs), and especially $k$-LIN, are the most fundamental optimization problems. An instance of a CSP consists of a set of $n$ variables and a set of $m$ local constraints where each constraint involves a small number of variables.  The goal is to decide if there exists an assignment to the variables that satisfies all the constraints. $3$-LIN is a special type of CSP where each constraint is a {\em linear equation} in the variables involved in the constraint. More specifically, a $3$-LIN instance over a (non-abelian or abelian) group $G$ has the constraints of the form  $a_{1} \cdot x_{1}\cdot a_{2} \cdot x_{2}\cdot a_{3} \cdot x_{3} = b$, where $a_{1}, a_{2}, a_{3}, b$ are the group elements and $x_1, x_2, x_3$ are the variables. One can also sometimes allow inverses of the variables in the equations.

For most CSPs, the  decision version is $\NP$-complete. Therefore, from the algorithmic point of view, one can relax the goal to finding an assignment that satisfies as many constraints as possible. An $\alpha$-approximation algorithm for a Max-CSP is an algorithm that always returns a solution that satisfies at least $\alpha\cdot\OPT$ many constraints, where $\OPT$ is the maximum number of constraints that can be satisfied by  an assignment. 

The famous  PCP theorem~\cite{DBLP:conf/focs/AroraLMSS92, DBLP:conf/focs/AroraS92, FeigeGLSS96} shows that certain Max-CSPs are hard to approximate within a factor $c<1$. 
A seminal result of H{\aa}stad~\cite{DBLP:journals/jacm/Hastad01} gives optimal inapproximability results for many CSPs including Max-$k$-SAT, Max-$k$-LIN over abelian groups, Set Splitting, etc. Once we understand the opitmal {\em worst-case} complexity of a Max-CSP, it is interesting to understand how the complexity of the problem changes under certain restrictions on the instances. One such example of restrictions is the study of promise CSPs \cite{AustrinGH17, BulinKO19} in which it is guaranteed that a richer solution exists (e.g., a given graph has a proper $3$-coloring) and the goal is to find or even approximate a weaker solution (e.g., find a coloring with $10$ colors that maximizes the number of non-monochromatic edges). Another example is the study of complexity of Max-CSP when the instance is generated using a certain random process~\cite{Feige02, KothariMOW17}.

In this paper we study the complexity of Max-$3$-LIN when the underlying constraints $vs.$ variables graph is fixed. A factor graph of an instance is a bipartite graph between $\{C_1, C_2, \ldots, C_m\}$ and $\{x_i, x_2, \ldots, x_n\}$ where we connect $C_i$ to $x_j$ if the constraint $C_i$ depends on the variable $x_j$. An instance can be completely described by its factor graph and by specifying which predicate to use for each constraint. Feige and Jozeph~\cite{DBLP:conf/icalp/FeigeJ12} were interested in understanding the effect of the factor graph on the complexity of approximating CSPs. Namely, for a given CSP, is there a factor graph $G_n$ for each input length $n$ such that the CSP remains hard even after restricting its factor graph to $G_n$? They answered this positively by showing that there exists a family of factor graphs for Max-3-SAT such that it is $\NP$-hard to approximate within a factor of s $\frac{77}{80} + \eps$, for every $\eps>0$. They defined such a family of graphs as the {\em universal factor graphs}.

\begin{definition}
	A family of $(c, s)$-universal factor graphs for a Max-CSP is a family of factor graphs, one for each input length, such that given a Max-CSP instance restricted to the factor graphs in the family, it is $\NP$-hard to find a solution with  value at least $s$ even if it is guaranteed that there is a solution with value at least $c$.
\end{definition}

In a follow-up work by Jozeph~\cite{Jozeph14}, it was shown that there are universal factor graphs for every $\NP$-hard Boolean CSP and for every APX-hard Boolean Max-CSP. However, the inapproximability factors in these results are weaker than what was known in the standard setting. 

The result of Feige and Jozeph~\cite{DBLP:conf/icalp/FeigeJ12} was vastly improved recently by  Austrin, Brown{-}Cohen and H{\aa}stad~\cite{DBLP:conf/soda/AustrinBH21}. They gave optimal inapproximability results for many well-known CSPs including Max-k-SAT, Max-TSA\footnote{each constraint is a Tri-Sum-And constraint, i.e.,~of the form $x_1 +
x_2 + x_3 + x_4 \cdot x_5 = b (\text{mod } 2)$}, ``$(2+\eps)$-SAT" and any predicate supporting a pairwise independent subgroup. Namely, they show the existence of $(1,\frac{7}{8}+\eps)$-universal factor graphs for a Max-$3$-SAT, $(1-\eps,\frac{1}{2}+\eps)$-universal factor graphs for a Max-3-LIN over $\mathbb{Z}_2$, etc., for every $\eps>0$. The optimal universal factor graph  inapproximability of Max-3-LIN over any finite {\em abelian} groups was also shown in~\cite{DBLP:conf/soda/AustrinBH21}. 

In this paper, we investigate the existence of universal factor graphs for Max-3-LIN over finite {\em non-abelian} groups.  Engebretsen, Holmerin and Russell~\cite{DBLP:journals/tcs/EngebretsenHR04} showed optimal inapproximability of Max-3-LIN over non-abelian groups with {\em imperfect} completeness.\footnote{The instances in~\cite{DBLP:journals/tcs/EngebretsenHR04} involve inverses in the equations, e.g., they are of the form $a_{1} \cdot x_{1}\cdot a_{2} \cdot x_{2}^{-1}\cdot a_{3} \cdot x_{3} = b$. } Recently, the first author and Khot~\cite{DBLP:journals/eccc/BhangaleK20} showed optimal inapproximability of Max-3-LIN over non-abelian groups with {\em perfect} completeness. Our main theorems extend both these results by showing the existence of universal factor graphs with the same hardness factor.

In the imperfect completeness case, we show that it is $\NP$-hard to do better than the random assignment even if the factor graph is fixed,  thereby extending the result of~\cite{DBLP:journals/tcs/EngebretsenHR04} to the universal factor graph setting.

\begin{theorem} 
	\label{theorem:imperfectlin}
	For every $\eps>0$ and any finite non-abelian group $G$, there are $(1-\eps, \frac{1}{|G|}+\eps)$-universal factor graphs for Max-$3$-LIN over $G$.
\end{theorem}

For a non-abelian group $G$, we denote by $[G,G]$ a commutator subgroup of $G$, i.e.,~the subgroup generated by the elements $\{g^{-1}h^{-1}gh\mid g,h\in G\}$. The factor $\frac{1}{|[G,G]|}$ comes naturally in the results on approximating Max-$3$-LIN over $G$ in the perfect completeness situation. This is because of the fact that $G/[G,G]$ is an abelian group and this can be used in getting the $\frac{1}{|[G,G]|}$-approximation algorithm for Max-$3$-LIN. For a concrete example, consider the group $S_3$, the group of all permutations of a three-element set. The commutator subgroup of $S_3$ is $\{(), (1,2,3), (1,3,2)\}$ which is isomorphic to $\mathbb{Z}_3$. In this case, $S_3/\mathbb{Z}_3 \cong \mathbb{Z}_2$ which is an abelian group. More generally, a given Max-$3$-LIN instance $\phi$ over $G$ can be thought of as a Max-$3$-LIN over $G/[G,G]$ by replacing the group constant by its coset in $G/[G,G]$. If $\phi$ is satisfiable over $G$, then $\phi'$ is satisfiable over $G/[G,G]$. The satisfying assignment for $\phi'$ can be found in polynomial time as $\phi'$ is a collection of linear equations over an abelian group. To get the final assignment for $\phi$,  we can assign with each variable $x$ a random element from the coset assigned by the satisfying assignment of $\phi'$. It is easy to see that this random assignment satisfies each equation of $\phi$ with probability $\frac{1}{|[G,G]|}$.

 In the perfect completeness situation too, we get the optimal universal factor graph hardness for Max-$3$-LIN, matching the inapproximability threshold of~\cite{DBLP:journals/eccc/BhangaleK20}.
\begin{theorem}
		\label{theorem:perfectlin}
	For every $\eps>0$ and any finite non-abelian group $G$, there are $(1, \frac{1}{|[G,G]|}+\eps)$-universal factor graphs for Max-$3$-LIN over $G$.
\end{theorem}

The actual theorem statements are stronger that what are stated. Along with the universal factor graph hardness, the instances have the following additional structure. 

\begin{enumerate}
	\item In the imperfect completeness case, our hardness result from Theorem~\ref{theorem:imperfectlin} holds for constraints of the form  $x_{1}\cdot x_{2}\cdot x_{3} = g$ for some $g\in G$. In~\cite{DBLP:journals/tcs/EngebretsenHR04}, the constraints involve inverses of the variables as well as group constants on the left hand side, for example, the constraints can be of the form  $a_{1} \cdot x_{1}\cdot a_{2} \cdot x_{2}^{-1}\cdot a_{3} \cdot x_{3} = b$. 
	\item Similar to the above, our hardness result from Theorem~\ref{theorem:perfectlin} holds even if we restrict the constraints in the Max-$3$-LIN instance to the form $x_{1}\cdot x_{2}\cdot x_{3} = g$ for some $g\in G$. In comparison,  in~\cite{DBLP:journals/eccc/BhangaleK20}, the definition of a linear equation involves using constants on the left-hand side of the equations. 
\end{enumerate}

To sum it up, our results show that the exact `literal patterns' as well as the factor graphs are not the main reasons for the optimal $\NP$-hardness of the aforementioned results~\cite{DBLP:journals/tcs/EngebretsenHR04, DBLP:journals/eccc/BhangaleK20} on Max-3-LIN over finite non-abelian groups.

\subsection{Techniques}

In this section, we highlight the main differences between the previous works~\cite{DBLP:journals/tcs/EngebretsenHR04, DBLP:conf/soda/AustrinBH21,DBLP:journals/eccc/BhangaleK20} and this work.

A typical way of getting optimal inapproximability result is to start with a gap $(1, 1-\eps)$ $\NP$-hard instance of a Max-CSP and apply parallel repetition on it to create a $2$-CSP (a.k.a. Label Cover, see Definition~\ref{def:LC_UFG}) with arbitrarily large gap of $(1,\delta)$, for any constant $\delta>0$.  Each vertex on one side of the Label Cover corresponds to a subset of constraints of the initial Max-CSP instance.  In order to reduce a Label Cover to a given Max-CSP over smaller alphabet, one key component in the reduction is to use a long-code encoding of the labels (i.e.,~the assignments to the variables in the constraints associated with the vertex) in the Label Cover instance. A long code of a label $i\in [n]$ is given by the truth-table of a function $f: [q]^n \rightarrow [q]$ defined as $f(x) = x_i$. One of the main reasons for using long code is that its high redundancy allows one to implicitly check if the assignment satisfies the given set of constraints associated with a vertex using the operation called {\em folding}. 
Thus, the only thing to check is if the given encoding is indeed (close to) a long code encoding of a label and hence it is successful in getting many tight inapproximability results. 

One issue with the foldings that appeared before the work of~\cite{DBLP:conf/soda/AustrinBH21} is that the folding structure changes if we change the literal structure of the underlying constraints. Therefore, even if we start with a universal factor graph hard instance of  a gap $(1, 1-\eps)$ Max-CSP, different literal patterns give different constraints $vs.$ variables graphs for the final Max-CSP instances. Therefore, the reduction template is not enough to get the same factor graph.

To overcome this difficulty the {\em functional folding} was introduced in~\cite{DBLP:conf/soda/AustrinBH21}. This folding is 'weaker' than the previously used foldings in terms of decoding to a label that satisfies the underlying predicate in the Label Cover instance. However, they show that this type of folding can be used to get the tight inapproximability results. The key lemma that was proved in~\cite{DBLP:conf/soda/AustrinBH21} is that non-zero Fourier coefficients of such a folded function correspond to sets of assignments with an odd number of them satisfying the constraints. 

We cannot directly use the functional folding defined in~\cite{DBLP:conf/soda/AustrinBH21} for two main reasons. Firstly, the way functional folding was defined, the underlying group is always an abelian group. Secondly, the aforementioned main lemma from~\cite{DBLP:conf/soda/AustrinBH21} talks about the non-zero Fourier coefficient of the folded function when viewed as a function over an abelian group. Since our soundness analyses use Fourier analysis over non-abelian groups (similar to~\cite{DBLP:journals/tcs/EngebretsenHR04, DBLP:journals/eccc/BhangaleK20} ), we cannot directly use their folding. 

In this work, we define functional folding for functions $f: G^n \rightarrow G$, where $G$ is a finite non-abelian group (See Definition~\ref{def:functional_folding}). One of the main contributions of this work is to prove that non-zero Fourier coefficients (Fourier matrices to be precise) of such functionally folded functions also have properties that are sufficient to complete the soundness analysis. 

For the proof of Theorem~\ref{theorem:imperfectlin}, it is enough to show that for functionally folded functions, any non-zero Fourier coefficient of the function has at least one assignment that satisfies all the constraints of the Label Cover vertex. We prove this in Lemma~\ref{functional_folding_use_almost_sat}. This conclusion is analogous to the one in~\cite{DBLP:conf/soda/AustrinBH21}.  In order to make sure that we do not use inverses in the constraints as mentioned in the previous section, we follow the proof strategy of~\cite{DBLP:journals/eccc/BhangaleK20} in the imperfect completeness case. Compared to~\cite{DBLP:journals/eccc/BhangaleK20}, in the imperfect completeness case, one needs to control extra terms related to dimension $1$ representations of the group which necessitates a different approach for the soundness analysis. In particular, we use the independent noise to take care of all high-dimensional terms in the Fourier expansion of the test acceptance probability.

For proving Theorem~\ref{theorem:perfectlin}, we need a stronger conclusion on the non-zero Fourier coefficients of functionally folded functions. This is because the decoding strategy in~\cite{DBLP:journals/eccc/BhangaleK20} is highly non-standard; The reduction can only decode from a very {\em specific subset} of the list corresponding to non-zero Fourier coefficients.  Therefore, we need to show that for this type of non-zero Fourier coefficients, there exists an assignment from that subset satisfying the constraints of the Label Cover vertex. This is done in Lemma~\ref{non_zero_coefficient_perfect_completness_lemma}.

We also observe that for the soundness proof to work, one does not need to fold all the functions. We use this observation to conclude that the reduced instance of Max-$3$-LIN has constraints of the form $x_{1}\cdot x_{2}\cdot x_{3} = g$ for some $g\in G$. Since the factor graph is fixed, these instances are completely specified by specifying only one group element per constraint.

\subsection{Organization}
We start with preliminaries in Section~\ref{section:Preliminaries}. In Section~\ref{subsection:reptheory}, we give an overview of representation theory and Fourier analysis over non-abelian groups. In Section~\ref{subsection:3linandUFG}, we formally define the problem that we study and universal factor graphs. In Section~\ref{function_folding_section}, we define functional folding and prove two key lemmas that will be used in the main soundness analysis. 

In Section~\ref{section:imperfect_3lin}, we prove the existence of $(1-\eps,\frac{1}{|G|}+\eps)$-universal factor graphs for Max-$3$-LIN over a non-abelian group $G$ with imperfect completeness. 
Finally, in Section~\ref{section:perfect_3lin}, we prove the existence of $(1,\frac{1}{|[G,G]|}+\eps)$-universal factor graphs for Max-$3$-LIN over a non-abelian group $G$ with perfect completeness. 

\section{Preliminaries}
\label{section:Preliminaries}
\subsection{Representation Theory and Fourier analysis on non-Abelian groups}
\label{subsection:reptheory}
In this section we give a brief description of concepts from representation theory used in this work. We will not prove any claims in this section; instead we refer the reader interested in a more detailed treatise to~\cite{Serre1977,terras_1999}. 
\par Let us start by introducing a definition of representation.
\begin{definition}
	A representation $(V_{\rho},\rho)$ of a group $G$  is a group homomorphism $\rho \colon G \to \GL(V_{\rho})$, where $V_{\rho}$  is a complex vector space.
\end{definition}
For the sake of brevity throughout this article we will use only symbol $\rho$  to denote a representation, and we will assume that the vector space $V_{\rho}$  can be deduced from the context. We work with finite groups, and hence assume that $V_{\rho} = \mathbb{C}^{n}$, where $n \in \mathbb{N}$, and that $\GL(V_{\rho})$ is a space of invertible matrices. Furthermore, we always work with unitary representations.
\par 
Study of representations can be reduced to the study of irreducible representations. In order to introduce them, we first bring in the following definition.
\begin{definition}
	Let $\rho$  be a representation of a group $G$. A vector subspace $W \subseteq V_{\rho}$  is $G$-invariant if and only if  
	\begin{equation*}
		(\forall g \in G, \forall w\in W) \quad \rho(g) w \in W.
	\end{equation*}
\end{definition}
Observe that if $W$  is $G$-invariant then the restriction $\left. \rho\right|_{W}$ of $\rho$ to $W$ is a representation of $G$. We can now introduce irreducible representations.
\begin{definition}
	A representation $\rho$ of a group	$G$ is irreducible if $V_{\rho} \neq \emptyset$ and its only $G$-invariant subspaces are $\left \{0\right\}$  and $V_{\rho}$.
\end{definition}
We use ${\bf \textrm{\textbf{Irrep}}(G)}$ to denote the set of all irreducible representations of $G$  up to an isomorphism, where an isomorphism between representations is given by the following definition.
\begin{definition}
	Two representations $\rho$ and $\tau$ of a group $G$ are isomorphic if there is an invertible linear operator $\varphi \colon V_{\rho} \to V_{\tau}$  such that
	\begin{equation*}
		\varphi \circ \rho(g) = \tau(g) \circ \varphi, \quad \forall g \in G.
	\end{equation*}
	We write $\rho \cong \tau$ to denote that $\rho$  is isomorphic to $\tau$.
\end{definition}
Representation theory is used in this work because it is a natural language for expressing Fourier analysis of the space $L^{2}	(G) = \left \{ f: G \to \mathbb{C} \right\}$, which will be an ubiquitous tool in this work. We endow the space $L^{2}(G)$  with the scalar product defined as 
\begin{equation*}
	\langle f_1, f_2 \rangle_{L^{2}(G)} = \frac{1}{|G|}\sum_{g \in G} f_1(g) \overline{f_2(g)},
\end{equation*}
which induces a norm on $L^{2}(G)$ via
\begin{equation*}
	\|f\|_{L^{2}(G)} = 	\sqrt{ \langle f, f \rangle_{L^{2}(G)}} = \sqrt{ \frac{1}{|G|}\sum_{g \in G} |f(g)|^{2} },
\end{equation*}
We can now introduce the Fourier transform as follows.
\begin{definition}
	For $f \in L^{2}(G)$, the Fourier transform of $f$ is an element $\hat{f}$ of $\prod_{\rho \in \Irrep(G)} \End (V_{\rho})$ given by
	\begin{equation*}
		\hat{f} (\rho) = \E_{x \in G} \left[ f(x) \rho(x) \right].
	\end{equation*}
\end{definition}
In the definition above we use $\End(V_{\rho})$  to denote a set of endomorphisms of the vector space $V_{\rho}$. In particular, once we fix bases of $\left \{ V_{\rho}\right\}_{\rho \in \Irrep(G)}$, we can identify $\hat{f}$ with a matrix. Throughout this article we will consider the space of matrices of same dimension to be equipped by the scalar product defined as
\begin{equation}\label{scalar_product_definition}
	\langle A, B \rangle_{\End(V_{\rho})	}= \tr(AB^{*}).
\end{equation}
Note that if we consider $A$  and $B$  as operators on a vector space $V$, the definition of $\langle A,B \rangle_{\End(V)}$ does not change under unitary transformations of basis. Let us now state the Fourier inversion theorem which shows that the function is uniquely determined by its Fourier coefficients.
\begin{lemma}[Fourier inversion]
	For $f \in L^{2}(G)$  we have 
	\begin{equation*}
		f(x) = \sum_{\rho\in \Irrep(G)} \dim(\rho) \langle	\hat{f}(\rho), \rho(x) \rangle_{\End(V_{\rho})}.
	\end{equation*}
\end{lemma}
Plancherel's identity can be written in this setting as follows.
\begin{lemma} (Plancherel's identity)
	\begin{equation*}
		\langle f,g \rangle_ {L^{2}(G)} = \sum_{\rho \in \Irrep(G)} \dim(\rho) \langle \hat{f}(\rho), \hat{g}(\rho)\rangle_{\End(V_{\rho})}.
	\end{equation*}
\end{lemma}
A straightforward corollary of Plancherel's idenity is Parseval's identity, given in the following lemma.
\begin{lemma} (Parseval's identity)
	For $f \colon G \to \mathbb{C}$  we have 
	\begin{equation*}
		\|f(x)\|_{L^{2}(G)}^{2} = \sum_{\rho \in \Irrep(G)} \dim(\rho) \|\hat{f}(\rho)\|_{HS}^{2},
	\end{equation*}
	where $\|\cdot\|_{HS}$ 	is a norm induced by the scalar product (\ref{scalar_product_definition}), i.e.,~it is the Hilbert-Schmidt norm defined on a set of linear operators on $V_{\rho}$ by 
	\begin{equation*}
		\|A\|_{HS} = \sqrt{ \langle A, A \rangle } = \sqrt{ \tr(A A^{*} ) } = \sqrt{\sum_{ij} |A_{ij}|^{2}}.
	\end{equation*}
\end{lemma}
The following lemma characterizes the values of scalar products between matrix entries of two representations. In particular, it shows that the matrix entries of two representations are orthogonal, and that $L^{2}(G)$-norm of each entry of representation $\rho$  equals to $1/\dim(\rho)$.
\begin{lemma} \label{orthogonality}
	If $\rho$  and $\tau$  are two non-isomorphic irreducible representations of $G$  then for any $i,j,k,l$  we have 
	\begin{equation*}
		\langle \rho_{ij}, \tau_{k l} \rangle_{L^{2}(G)} = 0.
	\end{equation*}
	Furthermore,
	\begin{equation*}
		\langle \rho_{ij}, \rho_{k l} \rangle_{L^{2}(G)} =  \frac{\delta_{ik} \delta_{jl}}{\dim(\rho)},
	\end{equation*}
	where $\delta_{ij}$ is the Kronecker delta function.
\end{lemma}
By taking $\tau \cong 1 $ in the previous lemma	we obtain the following corollary.
\begin{lemma}\label{corollary_1_chi}
	Let $\rho \in \Irrep(G)\setminus\left \{1\right\}$. Then
	\begin{equation*}
		\sum_{g \in G} \rho(g) = 0.
	\end{equation*}
\end{lemma}
Let us now associate a character with each representation.
\begin{definition}
	The character $\chi_{\rho}\colon G \to \mathbb{C}$ of a representation $\rho \colon G \to \GL(V)$ is a function defined by
	\begin{equation*}
		\chi_{\rho}(g) = \tr(\rho(g)).
	\end{equation*}
\end{definition}
\par
Characters are orthogonal to each other as shown by the following lemma.
\begin{lemma}
	For $\rho,\tau \in \Irrep(G)$ we have
	\begin{equation*}
		\langle \chi_{\rho}(g),\chi_{\tau}(g)\rangle_{L^2(G)} = 
		\begin{cases}
			1, \quad \rho \cong \tau, \\
			0, \quad \textrm{otherwise.}
		\end{cases}
	\end{equation*}
\end{lemma}
Another nice identity that characters satisfy is given in the following lemma.
\begin{lemma}\label{dim_rho_lemma}
		\begin{equation*}
			\sum_{\rho\in \Irrep(G)} \dim(\rho)\chi_{\rho}(g)= \begin{cases}
				|G|, \quad \textrm{if } g = 1_G,\\
				0, \quad \textrm{otherwise.}
			\end{cases}
		\end{equation*}  
\end{lemma}
Taking $g=1_G$ in the previous lemma implies that $\sum_{\rho \in \Irrep(G)} \dim(\rho)^{2}=|G|$, and hence for every $\rho \in \Irrep(G)$ we have $\dim(\rho)\leq \sqrt{|G|}$. \par
In this article we will also encounter convolution of functions in $L^{2}(G)$, which is defined as follows.
\begin{definition}
	Given $f,g \in L^{2}(G)$, their convolution $f*g \in L^{2}(G)$ is defined as 
	\begin{equation*}
		f*g(x) = \E_{y \in G} \left[f(y)g(y^{-1}x)\right].
	\end{equation*}
\end{definition}
Fourier analysis interacts nicely with the convolution, as shown by the following lemma.
\begin{lemma}
	For $f,g \in L^{2}(G)$ we have
	\begin{equation*}
		\widehat{f * g} (\rho) = \hat{f}(\rho)\cdot \hat{g}(\rho).
	\end{equation*}
\end{lemma}
Given a group $G$  we can define the group $G^{n}$ as the set of $n$-tuples of elements of $G$ on which the group operation is performed coordinate-wise. It is of interest to study the structure of representations of $G^{n}$, particularly in relation to representations of $G$. In order to do so, let us first introduce direct sum and tensor product of representations.
\begin{definition}
	Let $\rho$  and $\tau$  be two representations of $G$. We define their direct sum $\rho \oplus \tau$ to be a representation of $G$  defined on vectors $(v,w) \in \rho_V \oplus \tau_V$ by
	\begin{equation*}
		\rho\oplus \tau (v,w) = \rho(v) \oplus \tau(w).
	\end{equation*}
\end{definition}
Observe that if a representation $\rho$  is not irreducible then there are $G$-invariant vector subspaces $V_1,V_2 \subseteq V_{\rho}$  such that $V_{\rho}$ is a direct sum of $V_1$  and $V_2$, i.e.,~$V_{\rho}=V_1 \oplus V_2$. Hence, by fixing a suitable basis of $V_{\rho}$ we can write $\rho$ as a block diagonal matrix with two blocks on the diagonal corresponding to $\left. \rho \right|_{V_1}$ and $ \left. \rho \right|_{V_2}$, i.e.,~we have that $\rho \cong \left. \rho \right|_{V_1} \oplus \left. \rho \right |_{V_2}$. Since we assumed that $\dim(\rho)<\infty$ it follows by the principle of induction that we can represent each $\rho$ as 
\begin{equation*}
	\rho \cong \oplus_{s=1}^{u} \tau^s,
\end{equation*}
where $\tau^{s} \in \Irrep(G)$. We now introduce the tensor product of representations.
\begin{definition}
	Let $\rho$ and $\tau$  be representations of $G$. Tensor product $\rho \otimes \tau$ of $\rho$ and $\tau$  is a representation of $V_{\rho}\otimes V_{\tau}$ defined by
	\begin{equation*}
		(\rho \times \tau) (u \otimes v) = \rho(g)(u) \otimes \tau(g)(v),
	\end{equation*}
	and extended to all vectors of $V_{\rho}\otimes V_{\tau}$ by linearity. 	
\end{definition}
Following lemma characterizes irreducible representations of $G^{n}$ in terms of irreducible representations of $G$. 
\begin{lemma} \label{tensor_repr_G_n}
	Let $G$ be a group and let $n \in \mathbb{N}$. All irreducible representations of the group	$G^{n}$ are given by
	\begin{equation*} 
		\Irrep(G^{n}) = \left \{ \otimes_{d=1}^{n}  \rho_d\mid  \rho_d \in  \Irrep(G) \right\}.
	\end{equation*}
	We will sometimes use $\rho=(\rho_1,\hdots,\rho_n)$ to denote that $\rho=\otimes_{d \in [n]} \rho_d$. Furthermore, we will use $|\rho|$ to denote the number of $\rho_d\not \cong 1$.
\end{lemma}
In our arguments we will work with a map $\pi\colon [R] \to [L]$, where $L,R \in \mathbb{N}$, and the notation $[n]$ for $n \in \mathbb{N}$ is used to mean $[n]=\left \{1,2,\hdots,n\right\}$.
Let us define with $\tilde{\pi}\colon G^{L} \to G^{R}$ the map given by
\begin{equation*}
	\tilde{\pi} (x_1,\hdots,x_L) = (y_1,\hdots,y_R), \quad \textrm{where } y_i=x_{\pi(i)}.
\end{equation*}
If $\alpha \in \Irrep(G^{R})$, then the map $\alpha \circ	\tilde{\pi}$ is a representation of $G^{L}$. Observe that $\alpha \circ	\tilde{\pi}$ might not be irreducible, but anyhow it can be decomposed into its irreducible components, i.e.,~we have that 
\begin{equation*}
	\alpha\circ \tilde{\pi} \cong \oplus_{s=1}^{u}  \tau^{s},
\end{equation*}
where $u\in \mathbb{N}$ and $\tau^{s} \in \Irrep(G^{L})$. In particular, with a suitable choice of basis the representation $\alpha\circ \tilde{\pi}$ can be realized as a block diagonal matrix $\oplus_{s=1}^{u} \tau^{s}$. Hence, let us fix this basis and identify $\alpha\circ \tilde{\pi}$ with a $\dim(\alpha)\times \dim(\alpha)$ matrix with $u$  diagonal blocks. To each $1 \leq i,j \leq \dim(\alpha)$, such that $(i,j)$ belongs to some of the blocks we can assign a representation $\tau^{s} \in \Irrep(G^{L})$  and $1 \leq i',j'\leq \dim(\tau^{s})$ such that $(i,j)$-th matrix entry of $\alpha\circ \tilde{\pi}$ corresponds to $(i',j')$-th matrix entry of $\tau^{s}$. Throughout this work we will use $Q(\alpha,i,j)$ to denote $(\tau^{s},i',j')$ obtained by this correspondence. 
\par 
		Furthermore,  for a fixed $\alpha \in \Irrep(G^{R})$ and $1 \leq i \leq \dim(\alpha),$ let us denote with $\B(\alpha,i)$ the set of all $j\in \left \{1,\hdots,\dim(\alpha)\right\}$ such that $(i,j)$-th matrix entry of  $\alpha\circ \tilde{\pi}$ belongs to some block in the block diagonal matrix $\oplus_{s=1}^{u} \tau^{s}$. In particular, $B(\alpha,i)$ denotes all column indices $j$  which have non-trivial entry in their $i$-th row of $\alpha\circ \tilde{\pi}$. Observe that $B(\alpha,i)$ remains the same if we switch the roles of columns and rows in the previous definition. In particular, $\B(\alpha,i)$ is also the set of all $j\in \left \{1,\hdots,\dim(\alpha)\right\}$ such that $(j,i)$-th matrix entry of  $\alpha\circ \tilde{\pi}$ belongs to some block in the block diagonal matrix $\oplus_{s=1}^{u} \tau^{s}$. 
\par
Since $\alpha \in \Irrep(G^{R})$, by Lemma~\ref{tensor_repr_G_n} we have that $\alpha=\otimes_{d=1}^{R}\alpha_d$ where $\alpha_d \in \Irrep(G)$. Similarly, the representation $\tau^{s} \in \Irrep(G^{L})$ from $ \alpha\circ \tilde{\pi} \cong \oplus_{s=1}^{u}  \tau^{s}$ can be represented as $\tau^{s}= \otimes_{d=1}^{L} \tau^{s}_d$ where	$\tau^{s}_d \in \Irrep(G)$. Since all $\tau^{s}_d$ are completely determined by $\alpha$ and $\pi$, it is of interest to study the relation between $\tau^{s}_d$ and $\alpha_d$. In particular, we prove the following lemma, which states that if $\ell \in [R]$ is such that $\alpha_{\ell} \not \cong 1$ and for every other $\ell' \in [R] \setminus\left \{\ell\right\}$ such that $\pi(\ell') = \pi(\ell)$ we have $\alpha_{\ell'}\cong 1$, then $\tau^{s}_{\pi(\ell)} \not \cong 1$ for each $s$. 
\begin{lemma} \label{relationship_between_fourier_coeff_lemma}
	Let $\alpha \in \Irrep(G^{R})$, $\alpha=\otimes_{d=1}^{R} \alpha_d$, and let $\pi\colon [R] \to [L]$. Consider the representation $\alpha\circ \tilde{\pi} \cong \oplus_{s=1}^{u} \tau^{s}$ of $G^{L}$ where each $\tau^{s} \in \Irrep(G^{L})$. Furthermore, assume that for $\ell \in [R]$ we have that $\alpha_{\ell} \not \cong 1$, and also that for every $\ell' \in [R] \setminus \left \{\ell\right\}$ such that $\pi(\ell)=\pi(\ell')$ we have that $\alpha_{\ell'}\cong 1$. Then if we represent each $\tau^{s}$ as a tensor product of representations of $G$, i.e.,~$\tau^{s}=\otimes_{d=1}^L\tau^{s}_d$, we have that $\tau^{s}_{\pi(\ell)} \not \cong 1$.
	\begin{proof}
		Consider any block $\tau^s$ in the block diagonal matrix $\alpha \circ \tilde{\pi} \cong \oplus_{s=1}^{u} \tau^{s}$, and let $1\leq i',j'\leq \dim(\tau^{s})$ be arbitrary. Let $1\leq i,j \leq \dim(\alpha)$ be such that $[\alpha \circ \tilde{\pi}]_{ij}$ corresponds to $[\tau^{s}]_{i'j'}$, i.e.,~$i,j$ are unique such that $Q(\alpha,i,j) = (\tau^{s},i',j')$. Since $\alpha=\otimes_{d=1}^{R} \alpha_d$, we have that $[\alpha]_{ij} = \prod_{d=1}^{R}[\alpha_{d}]_{d_id_j}$, where $1\leq d_i,d_j \leq \dim(\alpha_d)$. But then
		\begin{equation*}
			[\tau^{s}]_{i'j'}({\bf x}) = [\alpha\circ \tilde{\pi}]_{ij} ({\bf x}) = \prod_{d=1}^{R}[\alpha_{d}]_{d_id_j} (x_{\pi(d)}) = \left(\prod_{d \in [R] \setminus \left \{\ell\right\}}[\alpha_{d}]_{d_id_j} (x_{\pi(d)})\right) \cdot [\alpha_{\ell}]_{\ell_i\ell_j}(x_{\pi(\ell)}).
		\end{equation*}
		Now since each $d \in [R]\setminus\left \{\ell\right\}$ satisfies either $\pi(d) \neq \pi(\ell)$ or $\alpha_d \cong 1$, we have that $\prod_{d \in [R] \setminus \left \{\ell\right\}}(\alpha_{d})_{d_id_j} (x_{\pi(d)})$ does not depend on $x_{\pi(\ell)}$. But then since $\alpha_{\ell} \not \cong 1$ we have that $[\tau^{s}]_{i'j'}({\bf x})$ depends on $x_{\pi(\ell)}$ for at least one $i',j'$, and therefore we can't have $\tau^{s}_{\pi(\ell)} \cong 1$. Since the choice of $s$ was arbitrary the claim of the lemma follows.
	\end{proof}
\end{lemma}

In the proof of the previous lemma we used the fact that $[\tau^{s}]_{i'j'}(x)=[\alpha\circ \tilde{\pi}]_{ij} (x) = \prod_{d=1}^{R}[\alpha_{d}]_{d_id_j} (x_{\pi(d)})$ where $Q(\alpha,i,j) = (\tau^{s},i',j')$. Hence, if for $\ell \in [L]$ and each $\ell' \in [R]$ such that $\pi(\ell')=\ell$ we have that $\alpha_{\ell'} \cong	1$, then $\tau^{s}_{\ell} \cong 1$ where $\tau^{s}_{\ell} \in \Irrep(G)$ is a component of the tensor product $\tau^{s}=\otimes_{d=1}^{L}\tau^{s}_d$ . In particular, we have the following lemma.
\begin{lemma}\label{relationship_between_fourier_coeff_lemma2}
	Let $\alpha \in \Irrep(G^{R})$, $\alpha=\otimes_{d=1}^{R} \alpha_d$, and let $\pi\colon [R] \to [L]$. Consider the representation $\alpha\circ \tilde{\pi} \cong \oplus_{s=1}^{u} \tau^{s}$ of $G^{L}$ where each $\tau^{s} \in \Irrep(G^{L})$. Furthermore let $\ell \in [L]$ be such that for every $\ell' \in [R], \pi(\ell')=\ell,$ we have that $\alpha_{\ell'}\cong 1$. Then for each $s=1,\hdots,u$, the representation $\tau^{s} \in \Irrep(G^{L})$ does not not depend on its $\ell$-th argument, i.e.,~in the  decomposition $\tau^{s} = \otimes_{d=1}^{L} \tau^{s}_{d}$ we have $\tau^{s}_{\ell}\cong 1$.
\end{lemma}
Since the statements of the previous two lemmas are somewhat cumbersome, let us summarize them into a claim with a more elegant description. Before that, we introduce the following definition.

\begin{definition}\label{definition_relationship_between_fourier_coeff}
	Let $\alpha \in \Irrep(G^{R})$, $\alpha=\otimes_{d=1}^{R} \alpha_d$, and let $\pi \colon [R] \to [L]$. We define $T(\alpha)$ to be the set of all $d$ for which $\alpha_d$ is non-trivial, i.e.,~
	\begin{equation*}
		T(\alpha) =\left \{ d \in [R] \mid \alpha_d \not \cong 1\right\},
	\end{equation*}
		and let $T_{\pi}(\alpha)$ be the set of all $\pi(d)$ where $d \in T(\alpha)$, i.e.,
		\begin{equation*}
			T_{\pi}(\alpha) = 	\left \{\pi(d) \mid d \in T(\alpha)  \right\}.
		\end{equation*}
		Finally, let $T^{uq}_{\pi}(\alpha)$ be the set of all $d \in T_{\pi}(\alpha)$ that are mapped uniquely by $\pi$, i.e.,~let 
	\begin{equation*}
		T_{\pi}^{uq}(\alpha) = 	\left \{\pi(d) \mid d \in T(\alpha) \textrm{ and } (\forall d' \in T(\alpha)\setminus \left \{d\right\}) \pi(d')\neq \pi(d)   \right\}.
	\end{equation*}
\end{definition}
Now, Lemma~\ref{relationship_between_fourier_coeff_lemma} and Lemma~\ref{relationship_between_fourier_coeff_lemma2} can be summarized as follows.
\begin{lemma}\label{relationship_between_fourier_coeff_imperfect_completeness}
	Let $\alpha \in \Irrep(G^{R})$, $\alpha=\otimes_{d=1}^{R} \alpha_d$, and let $\pi\colon [R] \to [L]$. Consider the representation $\alpha\circ \tilde{\pi} \cong \oplus_{s=1}^{u} \tau^{s}$ of $G^{L}$ where each $\tau^{s} \in \Irrep(G^{L})$. Then for each $s$ we have that $T_{\pi}^{uq}(\alpha) \subseteq T(\tau^{s}) \subseteq T_{\pi}(\alpha)$.
\end{lemma}
In our proofs of hardness of Max-3-Lin with perfect completeness in the decomposition of a representation $\alpha=\otimes_{d=1}^{R}\alpha_d$ instead of distinguishing between $\alpha_d \cong 1$ and $\alpha_d \not \cong	1$, we will need to distinguish between the case when $\dim(\alpha_d)=1$ or $\dim(\alpha_d)>1$. For that reason, we introduce analogues to Definition~\ref{definition_relationship_between_fourier_coeff} and Lemma~\ref{relationship_between_fourier_coeff_imperfect_completeness} as follows. 
\par 

\begin{definition}\label{definition_relationship_between_fourier_coeff_perfect}
	Let $\alpha \in \Irrep(G^{R})$, $\alpha=\otimes_{d=1}^{R} \alpha_d$, and let $\pi \colon [R] \to [L]$. Let $\tilde{T}(\alpha)$ be the set of all $d$ such that $\dim(\alpha_d) \geq 2$,  i.e.,~
	\begin{equation*}
		\tilde{T}(\alpha) =\left \{ d \in [R] \mid \dim(\alpha_d) \geq 2 \right\},
	\end{equation*}
		and let $\tilde{T}_{\pi}(\alpha)$ be the set of all $\pi(d)$ where $d \in \tilde{T}(\alpha)$, i.e.,
		\begin{equation*}
			\tilde{T}_{\pi}(\alpha) = 	\left \{\pi(d) \mid d \in \tilde{T}(\alpha)  \right\}.
		\end{equation*}
		Finally, let $\tilde{T}^{uq}_{\pi}(\alpha)$ be the set of all $d \in \tilde{T}_{\pi}(\alpha)$ that are mapped uniquely by $\pi$, i.e.,~let 
	\begin{equation*}
		\tilde{T}_{\pi}^{uq}(\alpha) = 	\left \{\pi(d) \mid d \in \tilde{T}(\alpha) \textrm{ and } (\forall d' \in \tilde{T}(\alpha)\setminus \left \{d\right\}) \pi(d')\neq \pi(d)   \right\}.
	\end{equation*}
\end{definition}

\begin{lemma}\label{relationship_between_fourier_coeff_perfect_completeness}
	Let $\alpha \in \Irrep(G^{R})$, $\alpha=\otimes_{d=1}^{R} \alpha_d$, and let $\pi\colon [R] \to [L]$. Consider the representation $\alpha\circ \tilde{\pi} \cong \oplus_{s=1}^{u} \tau^{s}$ of $G^{L}$ where each $\tau^{s} \in \Irrep(G^{L})$. Then for each $s$ we have that $\tilde{T}_{\pi}^{uq}(\alpha) \subseteq \tilde{T}(\tau^{s}) \subseteq \tilde{T}_{\pi}(\alpha)$.
\end{lemma}
The proof of Lemma~\ref{relationship_between_fourier_coeff_perfect_completeness} is deferred to the appendix.
\subsection{Max-3-Lin and Universal Factor Graphs}
\label{subsection:3linandUFG}
We begin by introducing the Max-3-Lin problem over a group $G$.
\begin{definition}
	In the Max-3-Lin problem over a group $G$ input is given by $n$  variables $x_1,\hdots,x_n,$ taking values in $G$, and $m$ constraints, where $i$-th constraint is of the form 
	\begin{equation*}
	 x_{i_1}\cdot x_{i_2}\cdot x_{i_3}=c_i \quad \textrm{for some } i_1,i_2,i_3 \in [n] \textrm{ and } c_i \in G.
	\end{equation*}
\end{definition}
Note that in this definition we do not allow for constants between the variables, i.e.,~we do not allow equations of the form $x_{i_1} \cdot g_i \cdot x_{i_2} \cdot g_i'\cdot x_{i_3} = c_i,$ with $g_i,g_i' \in G$, which was not the case with the previous works~\cite{DBLP:journals/eccc/BhangaleK20,DBLP:journals/tcs/EngebretsenHR04} on hardness of Max-3-Lin over non-abelian groups. Furthermore, this definition does not allow for inverses in equations, for example the constraint $x_{i_1} x_{i_2}^{-1} x_{i_3}^{-1}=c_i$ is not allowed, while in~\cite{DBLP:journals/tcs/EngebretsenHR04} these equations appeared since the analysis of the soundness required certain functions to be skew-symmetric (i.e.,~check Lemma~15 and Lemma~23 from~\cite{DBLP:journals/tcs/EngebretsenHR04}). Since in this work we consider hardness results, our proofs will also imply the hardness of instances defined in the sense of~\cite{DBLP:journals/eccc/BhangaleK20,DBLP:journals/tcs/EngebretsenHR04}.
\par Another strengthening of the previous results considered in this work comes from assuming that the instances have universal factor graphs. In order to formally explain this strengthening, we need to introduce the notion of a factor graph of a constraint satisfaction problem. We first recall the definition of a constraint satisfaction problem.
\begin{definition}
	A constraint satisfaction problem (CSP) over a language $ \Sigma = [q], q \in \mathbb{N}$, is a finite collection of predicates $K \subseteq \{ P:[q] ^k \to \{0,1\} \mid k \in \mathbb{N}\}$. 
\end{definition}
We use use $\ar(P)=k$ to denote the arity of a predicate $P:[q] ^k \to \{0,1\}$. As an input to our problem we get an instance of a CSP $K$, which is defined as follows.
\begin{definition} \label{csp_inst_def}
	An instance $\mathcal{I}$ of a CSP $K$ consists of a set $X = \{x_1,\hdots,x_n\}$ of $n$ variables taking values in $\Sigma$ and  $m$ constraints $C_1,\hdots,C_{m}$, where each constraint $C_i$ is a pair $(P_i,S_i)$, with $P_i \in K$ being a predicate with arity $k_i := \ar(P_i)$, and $S_i$ being an ordered tuple containing $k_i$ \emph{distinct} variables which we call scope.
\end{definition}
Let us denote by $\sigma \colon X  \to \Sigma$ an assignment to the variables $X$ of a CSP instance $\mathcal{I}$. We interpret $\sigma(S_i)$ as a coordinate-wise action of $\sigma$ on $S_i$. Given $\sigma$, we define the value $\Val_{\sigma}(\mathcal{I})$ of $\sigma$ as 
\begin{equation}
	\Val_{\sigma}(\mathcal{I}) = \sum\limits_{r=1}^m P_r(\sigma(S_r)).
\end{equation}
We work with Max-CSPs in which we are interested in maximizing $\Val_{\sigma}(\mathcal{I})$. We use	$\Opt(\mathcal{I})$ to denote the optimal value of $\mathcal{I}$, i.e.,~$\Opt(\mathcal{I})  = \max\limits_{\sigma} \left( \Val_{\sigma}(\mathcal{I}) \right).$ Observe that Max-3-Lin over $G$ can be seen as a Max-CSP with predicates  $K=\left \{P_g\right\}_{g \in G}$, where 
\begin{equation*}
	P_g( x,y,z)= 
	\begin{cases}
		1,& \quad \textrm{if }		x\cdot y \cdot z = g,\\
		0,& \textrm{otherwise}.
	\end{cases}
\end{equation*}
We can now introduce factor graphs and the notion of hardness we are interested in this article.
\begin{definition}
	The factor graph $\mathcal{F}$ of an instance $\mathcal{I}$ consists of the scopes $S_r,r=1,\hdots,m$. A family $\left \{\mathcal{F}_n\right\}_{n \in \mathbb{N}}$ is explicit if it can be constructed in time that is polynomial in $n$.
\end{definition}
\begin{definition}
	We say that the Max-CSP($K$) is  $(c,s)$-UFG-NP-hard if there is an explicit family of factor graphs $\left \{\mathcal{F}_n\right\}_n$  for $K$ and a polynomial time reduction $R$ from 3-Sat instances $I_n$ on $n$ variables to Max-CSP($K$) instances $R(\mathcal{I}_n)$ with the factor graph $\mathcal{F}_n$ such that the following holds
	\begin{itemize}
		\item \emph{Completeness:} If $I_n$ is satisfiable then $\Opt(R(\mathcal{I})) \geq c$.
		\item \emph{Soundness:} If $I_n$ is not satisfiable then $\Opt(R(\mathcal{I})) \leq s$.
	\end{itemize}
	We will say that Max-CSP($K$) has ``universal factor graphs'' to mean that Max-CSP($K$) is $(c,s)$-UFG-NP-hard, in which case the values of $c$ and $s$ will be clear from the context. 
\end{definition}
\begin{definition}
	A polynomial time reduction R from CSP $K$ to CSP $K'$ is factor graph preserving if it satisfies the following property:
	\begin{itemize}
		\item	Whenever two instances $I_1,I_2$  of $K$ have the same factor graphs, the respective instances $R(I_1)$ and $R(I_2)$ output by the reduction $R$  have the same factor graphs as well.
	\end{itemize}
\end{definition}
The definition of the $(c,s)$-UFG-NP-hardness given here matches the one from~\cite{DBLP:conf/soda/AustrinBH21}. Note that another view is given by the definition from~\cite{DBLP:conf/icalp/FeigeJ12} which considered hardness from the perspective of circuit complexity, i.e.,~with the starting point being $\NP \not \subseteq \textrm{P}/\textrm{Poly}$ one would aim to show non-existence of polynomially sized circuits which distinguish soundness from completeness of instances with fixed factor graphs. This difference is only technical, and in particular it is not hard to see that the analogues of all the results given in this work also hold in the alternative setting considered in~\cite{DBLP:conf/icalp/FeigeJ12}.
\par 
Majority of the strong hardness of approximation reductions such as~\cite{DBLP:journals/jacm/Chan16,DBLP:journals/jacm/Hastad01} use as an intermediary step in a reduction the parallel repetition~\cite{DBLP:journals/siamcomp/Rao11,DBLP:journals/siamcomp/Raz98} of 3-Sat instances output by the PCP theorem~\cite{DBLP:conf/focs/AroraLMSS92,DBLP:conf/focs/AroraS92,DBLP:journals/jacm/Dinur07}. Let us briefly describe the parallel repetition as a game played between two provers that can not communicate with each other and a verifier. We first fix the number of repetitions $r \in \mathbb{N}$. Then, in each round the verifier picks $r$  random constraints $C_{i_1},C_{i_2},\hdots,C_{i_r}, 1 \leq i_1,\hdots,i_r \leq m,$ where $C_{i_j}$ are constraints of the starting 3-Sat instance with $m$ constraints. Then, the verifier sends these $r$ constraints to the prover $P_1$. The verifier also picks one variable from each constraint sent to $P_1$ and sends these variables to the prover $P_2$. The first prover responds with a satisfying assignment to the $r$ constraints, while $P_2$  responds with values to $r$  variables. The verifier accepts if the assignments of the provers $P_1$  and $P_2$  agree on the same variables. \par
Parallel repetition game is usually conceptualized by the Label Cover problem, whose definition can be given as follows.
\begin{definition}
	A Label Cover instance $\Lambda=\left(V,W,E,[L],[R],\Pi\right)$ is a tuple in which 
	\begin{itemize}
		\item $(V,W,E)$	is a bipartite graph with vertex sets $V$ and $W$, and an edge set $E$.
		\item $[L]$ and $[R]$ are alphabets, where $L,R \in \mathbb{N}$.
		\item $\Pi$ is a set which consists of projection constraints $\pi_e: [R] \to [L]$ for each edge  $e \in E $.
	\end{itemize}
	The value of a Label Cover instance under an assignment $\sigma_L \colon V \to [L]$, $\sigma_R \colon W \to [R],$ is defined as the fraction of edges $e \in E$  that are satisfied, where an edge $e=(v,w)$ is satisfied if $\pi_e(\sigma_R(w)) = \sigma_L(v)$. We will write $\Val_{\sigma}(\Lambda)$  for the value of the Label Cover instance $\Lambda$ under an assignment $\sigma=(\sigma_L,\sigma_R)$.
\end{definition}
In this definition one can see $[R]$ as a set of all possible satisfying assignments of the prover $P_1$  for a question $w \in W$. Hence, the set $[R]$ depends not only on the factor graph of the starting 3-Sat instance but also on the predicates applied to each triplet of variables. Due to this obstacle we can not use Label Cover as the black box input to our problem. In particular, we need to resort to bookkeeping the types of predicates received by the prover $P_1$. Let us now describe the approach taken in this work.
\par The first difference compared to the ``classical'' approach to parallel repetition described above comes from the fact that as in~\cite{DBLP:conf/soda/AustrinBH21} we apply parallel repetition to Max-TSA problem which is $(1,1-\varepsilon)$-UFG-NP-hard~\cite{DBLP:conf/icalp/FeigeJ12} for some $\varepsilon>0$. Max-TSA problem is a CSP with predicates $TSA_b\colon \left \{0,1\right\}^{5}\to \left \{0,1\right\}, b \in \left \{0,1\right\},$  given by
\begin{equation*}
	TSA_b	(x_1,x_2,x_3,x_4,x_5) =
	\begin{cases}
		1, & \quad \textrm{ if }x_1 + x_2 + x_3 +x_4 \cdot x_5=b,\\
		0, & \textrm{ otherwise}.
	\end{cases}
\end{equation*} 
The reason for using Max-TSA problem instead of 3-Sat is technical in nature and it has to do with the fact in the case we use Max-TSA and we identify with $[R]$ all the possible assignments to the $5r$ variables received by $P_1$, then checking whether an assignment $\ell \in [R]$ returned by the prover satisfies all the constraints is equivalent to checking whether $r$ equations 
\begin{equation*}
	\mathscr{p}_j(\ell)=b_{i_j}, \quad j \in [r],
\end{equation*}
are true. This is achieved by setting each $\mathscr{p}_j$ to be the function that extracts values of variables appearing in $j$-th constraint received by $P_1$, and if the values of the variables are $x_1,x_2,\hdots,x_5$, then the function $\mathscr{p}_j$ returns the value $x_1 + x_2 + x_3 +x_4 \cdot x_5$. Observe that the parallel repetition applied to 3-Sat instances does not admit this description, in particular since for 3-Sat a single assignment to a triplet of variables can satisfy more than one 3-Sat predicate. This difference turns out to be crucial for the functional folding which will be introduced in Section~\ref{function_folding_section}. 
\par
We can now conceptualize the game between the two provers described above with the following definition.
\begin{definition}
	\label{def:LC_UFG}
	A UFG Label Cover instance $\Lambda_{UFG}$ is defined as $\Lambda_{UFG}=(\Lambda, U)$, where $\Lambda=\left(V,W,E,[L],[R],\Pi\right)$  is a Label Cover instance, while $U=\left( \mathcal{P}, {\bf b}, I \right)$ consists of
	\begin{itemize}
		\item $\mathcal{P}$, which is a set of functions $\mathscr{p}_1,\hdots,\mathscr{p}_r \colon [R] \to \left \{0,1\right\}$,  and
		\item ${\bf b} \in \left \{0,1\right\}^{m}, {\bf b}=(b_1,\hdots,b_m)$, 
		\item $I$, which is a collection of functions $I_w: [r] \to [m], w \in W$.
	\end{itemize}
	The value of a UFG Label Cover instance under an assignment $\sigma=(\sigma_L,\sigma_R)$  is the fraction of satisfied edges of $\Lambda$, where an edge $e=(w,v)$ is satisfied if it satisfies the projective constraint and if furthermore $\mathscr{p}_i(\sigma_R(w)) = b_{I_w(i)}$ for $i=1,\hdots,r$.
\end{definition}
In the definition above $I_w$ returns the indices of the constraints received by the prover $P_1$ when given a query $w$. \par
Since the starting Max-TSA problem is $(1,0.51)$-UFG-NP-hard~\cite{DBLP:conf/soda/AustrinBH21}, by the parallel repetition theorem~\cite{DBLP:journals/siamcomp/Raz98} we have the following lemma.
\begin{lemma}
	For any $\gamma>0$ there is a reduction from $(1,0.51)$-Max-TSA to UFG Label Cover $\Lambda_{UFG}(\Lambda,U)$ such that it is NP-hard to distinguish the following cases 
	\begin{itemize}
		\item \emph{Completeness}	$\Opt(\Lambda_{UFG})=1$,
		\item \emph{Soundness} $\Opt(\Lambda_{UFG}) < \gamma$.
	\end{itemize} Furthermore, starting from two instances of Max-TSA with the same factor graphs, the reduction will produce instances $\Lambda_{UFG},\Lambda_{UFG}'$ which differ only in their respective values of vectors ${\bf b},{\bf b'}$.
\end{lemma}
In our work for technical reasons we actually use smooth parallel repetition, which we describe in terms of two-prover game as follows. In this version apart from $r$  we fix $t \in \mathbb{N}, 1 < t <r$ as well, and instead of picking $r$  variables from the constraints $C_{i_1},\hdots,C_{i_r}$ sent to $P_1$ and sending them to $P_2$, the verifier selects $t$  constraints at random and sends one variable from each of these constraints to $P_2$. The verifier also sends $r-t$  remaining constraints to $P_2$. Note that in this version we do not need to force $P_2$  to satisfy these constraints since this is enforced by the agreement test performed by the verifier. We use smooth parallel repetition because we want to say that for any two answers given by $P_1$  it is highly likely that $P_2$ will have only one answer that will be accepted by the verifier. UFG Label Cover described in this paragraph will be referred to as $(r,t)$-smooth UFG Label Cover. 
\par 
In order to express the smoothness property of $\Lambda_{UFG}$ we introduce the following definition.
\begin{definition}
	Consider $\pi\colon [R] \to [L]$ and let $S \subseteq [R]$. We use $\mathcal{C}(S,\pi)$  to denote the indicator variable which is equal to $1$ if and only if there are two distinct $i,j \in S$ such that $\pi(i)=\pi(j)$, or formally:
	\begin{equation*}
		\mathcal{C}(S,\pi) = \begin{cases}
			1, \quad \textrm{if } (\exists i,j \in S, i \neq j), \pi(i) = \pi(j),\\
			0, \quad \textrm{otherwise}.
		\end{cases}
	\end{equation*}
\end{definition}
Now, as shown by Claim 2.20 in~\cite{DBLP:conf/soda/AustrinBH21}, we can assume that the UFG Label Cover instance satisfies \emph{smoothness} property defined as follows.
\begin{lemma}\label{smoothness_lemma}
	Let $0 \leq t \leq r, t \in \mathbb{N}$, consider $(r,t)$-smooth UFG Label Cover $\Lambda_{UFG}$ and let $w \in W$ be any vertex of $W$. If we denote with $E_w$ the set of of all edges $e$ incident to $w$, then for $S \subseteq [R]$ we have that
\begin{equation*}
	\E_{e \in E_w} \left[ \mathcal{C}(S,\pi_e) \right] \leq \frac{|S|^{2} t }{r}.
\end{equation*}
\end{lemma}
The difference in the lemma above and Claim 2.20 in~\cite{DBLP:conf/soda/AustrinBH21} is due to the fact that we use $r$ and $t$ here in the role played by $(r+t)$ and $r$ in~\cite{DBLP:conf/soda/AustrinBH21}, respectively. Let us also observe that choosing sufficiently large $t$ ensures that the parallel repetition of $(1,0.51)$-UFG-NP-hard Max-TSA still yields $\Lambda_{UFG}$ instance with completeness $1$  and soundness $\gamma$.

\subsection{Functional Folding Meets Representation Theory}\label{function_folding_section}
The next step in the ``classical reduction'' consists in encoding the answers of provers to a query $w$ for $P_1$ (or $v$  for $P_2$)   by a long code which can be thought of as a function $f_w\colon G^{R}\to G$ for $P_1$ (or $f_v \colon G^{L} \to G$ for $P_2$), and using a suitable test along each edge of a Label Cover instance that depends on the hardness result we are trying to prove. Typically, in order to avoid degenerate cases, we require functions $f_w$  and $f_v$ to be folded by requiring them to satisfy
\begin{equation*}
	f_w(c{\bf x}) = c f_w({\bf x}), \quad f_v(c{\bf x}) = cf_v({\bf x}), 
\end{equation*}
where $c \in G$ and for	${\bf x}=(x_1,\hdots,x_R)$ the action of $c$ on ${\bf x}$  is performed coordinate-wise, i.e.,~$c{\bf x} = (cx_1,\hdots,cx_R)$. However, this folding does not depend on the values of $b_i$ received by the prover $P_1$, and hence the provers can come up with a strategy for only one ${\bf b}$ which invalidates desired soundness property of the reduction. For that reason we use here functional folding, which was the technical novelty introduced in~\cite{DBLP:conf/soda/AustrinBH21}. 
\begin{definition}
	\label{def:functional_folding}
	Given a fixed collection of functions $\mathscr{p}_1,\mathscr{p}_2,\hdots,\mathscr{p}_r \colon [R] \to \left \{0,1\right\}$, let us introduce an equivalence relation $\sim$ on $G^{R}$ by
	\begin{equation*}
		{\bf x} \sim {\bf y} \iff (\exists F\colon \left \{0,1\right\}^{r} \to G) \textrm{ such that } \left(\forall d \in [R]\right) x_d  = F(\mathscr{p}_1(d),\mathscr{p}_2(d),\hdots,\mathscr{p}_r(d)) y_d.
	\end{equation*}
	Let ${\bf b}\in \left \{0,1\right\}^{r}, {\bf b} = (b_1,\hdots,b_r)$. We say that a function $f_{\bf b} \colon G^{R} \to G$ is functionally folded with respect to $(\left \{\mathscr{p}_i\right\}_{i=1}^{r},\textbf{b})$ if for every ${\bf x} \sim {\bf y}$ such that 
	\begin{equation*}
		\left(\forall d \in [R]\right) x_d  = F(\mathscr{p}_1(d),\mathscr{p}_2(d),\hdots,\mathscr{p}_r(d)) y_d, \quad \textrm{for some }F\colon \left \{0,1\right\}^{r} \to G,
	\end{equation*}
	we have 
	\begin{equation}\label{folding_relation}
		f_{\bf b}({\bf x}) = F(b_1,b_2,\hdots,b_r) f_{\bf b}({\bf y}).
	\end{equation}
	For the sake of notational convenience we will usually omit the subscript ${\bf b}$ and also not mention that the folding is with respect to $(\left \{\mathscr{p}_i\right\}_{i=1}^{r},{\bf b})$, especially when this can be easily inferred from the context.
\end{definition}
Note that in order to construct a functionally folded function $f$ we only need to define it on a fixed representative ${\bf y}$ of each equivalence class $[{\bf y}]$. We can then extend the value of $f$ to any ${\bf x} \sim {\bf y}$ by saying that 
\begin{equation*}
	f({\bf x}) = 	F(b_1,b_2,\hdots,b_r) f({\bf y}).
\end{equation*}
As an immediate consequence of the definition we have the following lemma.
\begin{lemma}
	If $f \colon G^{R} \to G$ is functionally folded, then for any $c \in G$  and any	${\bf x} \in G^{n}$  we have 
	\begin{equation*}
		f(c{\bf x}) = c f({\bf x}).
	\end{equation*}
	\begin{proof}
		By choosing $F \equiv c$ we have that $c{\bf x} \sim {\bf x}$, and since $f$  is folded by (\ref{folding_relation}) we have $f(c{\bf x}) = c f({\bf x})$, irrespective of the values of $b_i$  since $F$  is constant.
	\end{proof}	
\end{lemma}
Following lemma allows us to eliminate trivial representations in the Fourier spectrum of a function $\rho\circ f$, where $f$  is functionally folded.
\begin{lemma} \label{trivial_fourier_0_folding_lemma}
	Let $f \colon G^{n} \to G$ satisfy $f(c{\bf x}) = cf({\bf x})$ for every $c \in G$  and every ${\bf x} \in G^{n}$, and let $\rho \in \Irrep(G) \setminus \left \{1\right\}$. Then for $g=[\rho\circ f]_{pq}$, where $1 \leq p,q \leq \dim(\rho)$,  if $\alpha\cong	1$ we have $\hat{g}(\alpha)=0$. The same statement holds if we replace the condition that $f$  satisfies $f(c{\bf x})=cf({\bf x})$ by $f({\bf x}c) = f({\bf x}) c$  for every $c \in G$.
	\begin{proof}
		\begin{equation*}
			\begin{split}
				\hat{g}(1) & = \E_{{\bf x} \in G^{n}} \left[ g({\bf x}) \cdot 1 \right]= \E_{{\bf x} \in G^{n}} \left[ \rho(f({\bf x}))_{pq} \right]= \frac{1}{|G|} \E_{{\bf x} \in G^{n}} \left[ \sum_{c \in G}\rho(f(c{\bf x}))_{pq} \right] \\
						 & =\frac{1}{|G|} \E_{{\bf x} \in G^{n}} \left[ \sum_{c \in G} \sum_{1 \leq r \leq \dim(\rho)} \rho(c)_{pr} \rho(f({\bf x}))_{rq} \right] =\frac{1}{|G|} \sum_{1 \leq r \leq \dim(\rho)}\E_{{\bf x} \in G^{n}} \left[  \rho(f({\bf x}))_{rq}\right]  \sum_{c \in G}\rho(c)_{pr} =0,
			\end{split}
		\end{equation*}
		where the last inequality holds since $\sum_{c \in G}\rho(c)_{pr}=0 $  by Lemma~\ref{corollary_1_chi}. The proof in case $f({\bf x} c) = f({\bf x}) c$ is analogous.
	\end{proof}
\end{lemma}
The lemma above will be useful when deriving the result for Max-3-Lin with almost perfect completeness. For Max-3-Lin over $G$  with perfect completeness we will actually view all representations $\beta \in \Irrep(G^{L})$ with $\dim(\beta)=1$ as trivial, and hence we will need the following lemma.
\begin{lemma} \label{trivial_fourier_0_folding_lemma_perfect_completeness}
	Let $f \colon G^{n} \to G$ satisfy $f(c{\bf x}) = cf({\bf x})$ for every $c \in G$  and every ${\bf x} \in G^{n}$, and let $\rho \in \Irrep(G), \dim(\rho) \geq 2$. Then for $g=[\rho\circ f]_{pq}$ where $1\leq p,q \leq \dim(\rho),$ and  $\alpha \in \Irrep(G^{n}), \dim(\alpha)=1,$ we have that $\hat{g}(\alpha)=0$. The same statement holds if we replace the condition that $f$  satisfies $f(c{\bf x})=cf({\bf x})$ by $f({\bf x}c) = f({\bf x}) c$  for every $c \in G$.
\end{lemma}
The statement and the proof of this lemma already appeared in~\cite{DBLP:journals/eccc/BhangaleK20} as Lemma 2.29, and hence we omit the details here. \par
Observe that the proof of Lemma~\ref{trivial_fourier_0_folding_lemma} and Lemma~\ref{trivial_fourier_0_folding_lemma_perfect_completeness} did not require any additional properties of functional folding, and therefore they can be applied for $f$ that is ``classically folded'' as well. 
\par The next two lemmas we prove can be thought of as strengthenings of Lemma~\ref{trivial_fourier_0_folding_lemma} and Lemma~\ref{trivial_fourier_0_folding_lemma_perfect_completeness}, respectively, for functionally folded functions $f$. In particular they show that if $f$  is functionally folded, then for $g=[\rho\circ f]_{pq}$  and $\alpha\in \Irrep(G^{R}), \alpha=(\alpha_1,\hdots,\alpha_R)$  such that $\hat{g}(\alpha)\neq 0$ we need to have a ``non-trivial'' $\alpha_{\ell}$ such that $\ell \in [R]$ satisfies all constraints $\mathscr{p}_i(\ell)=b_i$. \par
We first state and prove the lemma that will be useful for the soundness analysis of Max-3-Lin with almost perfect completeness. 
\begin{lemma}\label{functional_folding_use_almost_sat}
	Let $f \colon G^{R} \to G$  be functionally folded with respect to $(\left \{\mathscr{p}_i\right\}_{i=1}^{r},{\bf b})$, and let $\rho \in \Irrep(G)$  be a representation such that $\rho \not \cong 1$. Given some $1 \leq p,q \leq \dim(\rho)$, let $g({\bf x}) = \rho(f({\bf x}))_{pq}$, and consider $\alpha \in \Irrep(G^{R})$, $\alpha=(\alpha_1,\hdots,\alpha_R)$. If for every $d \in [R]$ such that $\alpha_{d} \not \cong 1$ we have $(\mathscr{p}_1(d),\hdots,\mathscr{p}_r(d))\neq {\bf b}$, then $\hat{g}(\alpha) = 0$.
	\begin{proof}
		For $c \in G $ let us define a function $F_c \colon \left \{0,1\right\}^{r} \to G$ by
		\begin{equation*}
			\begin{cases}
				F_c({ \bf t})=c, \quad \textrm{if } {\bf t}={\bf b},\\
				F_c({ \bf t}) =1_G, \quad \textrm{otherwise},
			\end{cases}
		\end{equation*}
		and let us denote with $\vec{F_c}$ the vector in $G^{R}$  defined by $(\vec{F_c})_d = F_c(\mathscr{p}_1(d),\hdots,\mathscr{p}_r(d)), 1 \leq d \leq R$. Then for $1 \leq i,j \leq \dim(\alpha)$ we can write
		\begin{equation*}
			\hat{g}(\alpha)_{ij}= \E_{{\bf x} \in G^{R}}\left[ g({\bf x}) \alpha({\bf x})_{ij} \right] = \frac{1}{|G|}\E_{{\bf x} \in G^{R}}\left[\sum_{c \in G} g(\vec{F_c} {\bf x}) \alpha(\vec{F_c}{\bf x})_{ij} \right] .
		\end{equation*}
		Let us now fix ${\bf x} \in G^{R}$ 	and study the term $\sum_{c \in G} g(\vec{F_c} {\bf x}) \alpha(\vec{F_c}{\bf x})_{ij}$. Denote with ${\bf y}^{c}:=\vec{F_c}{\bf x}$, and observe that for each $c \in G$ we have  ${\bf y}^{c} \sim {\bf x}$ since
		\begin{equation*}
			y^{c}_d = F_c(\mathscr{p}_1(d),\hdots,\mathscr{p}_r(d)) x_d, \quad (\forall d \in [R]).
		\end{equation*}
			Therefore, since $f$ is functionally folded we have that
			\begin{equation*}
				g(\vec{F_c} {\bf x} ) = \rho(f(\vec{F_c}{\bf x}))_{pq} =\rho(c(f({\bf x}))_{pq}.
			\end{equation*}
			Let us now study the term $\alpha(\vec{F_c}{\bf x})_{ij}$. Using Lemma~\ref{tensor_repr_G_n} let us represent $\alpha$ as the tensor product of irreducible representations $\alpha_d \in \Irrep(G), d=1,\hdots,R$, i.e.,~let us write $\alpha({\bf x})=\alpha_1(x_1) \otimes \alpha_2(x_2) \otimes \hdots\otimes \alpha_R(x_R)$. Consider any $d \in [R]$ such that $\alpha_{d} \not \cong  1$. By the assumption of the lemma we have $(\mathscr{p}_1(d),\hdots,\mathscr{p}_r(d)) \neq {\bf b}$. But in this case $F_c(\mathscr{p}_1(d),\hdots,\mathscr{p}_r(d))=1_g$ and since 
			\begin{equation*}
				y_{d}^c = F_c(\mathscr{p}_1(d),\hdots,\mathscr{p}_r(d)) x_{d},
			\end{equation*}
			we have that $x_{d}=y_{d}^c$. Since for the remaining $d$ we have that $\alpha_d$ are constant (i.e.,~$\alpha_d \cong 1$), we have that $\alpha(\vec{F_c}{\bf x})_{ij} = \alpha({\bf x})_{ij}$. Hence we can write
			 \begin{equation*}
				 \hat{g}(\alpha)_{ij}	 = \frac{1}{|G|}\E_{{\bf x} \in G^{R}}\left[\sum_{c \in G} \rho(cf({\bf x}))_{pq} \alpha({\bf x})_{ij} \right] . 
			 \end{equation*}
			 Finally, we have that
			 \begin{equation*}
				 \sum_{c \in G} \rho(cf({\bf x}))_{pq}  = \sum_{c \in G} (\rho(c) \rho(f({\bf x})))_{pq} = \sum_{c \in G} \sum_{1\leq r \leq \dim(\rho)} \rho(c)_{pr}\rho(f({\bf x}))_{rq} =0,
			 \end{equation*}
			 where we used the fact that $\sum_{c \in G} \rho(c)_{pr}=0$, which holds since $\rho \not \cong 1$ allows us to use Lemma~\ref{orthogonality}. 
	\end{proof}
\end{lemma}
The next lemma we prove will be useful for decoding strategies of the provers in the analysis of soundness for Max-k-Lin with perfect completeness.
\begin{lemma}\label{non_zero_coefficient_perfect_completness_lemma}
	Let $f \colon G^{R} \to G$  be functionally folded  with respect to $(\left \{\mathscr{p}_i\right\}_{i=1}^{r},{\bf b})$, and let $\rho \in \Irrep(G)$  be a representation such that $\dim(\rho)\geq 2$. Given some $1 \leq p,q \leq \dim(\rho)$, let $g({\bf x}) = \rho(f({\bf x}))_{pq}$, and consider $\alpha\in \Irrep(G^{R})$, $\alpha=(\alpha_1,\hdots,\alpha_R)$. If for each $d \in [R]$ such that $\dim(\alpha_d)\geq 2$ we have that $(\mathscr{p}_1(d),\hdots,\mathscr{p}_r(d)) \neq {\bf b}$, then $\hat{g}(\alpha) = 0$.
	\begin{proof}
		For $c \in G $ we let a function $F_c \colon \left \{0,1\right\}^{r} \to G$ and a vector $\vec{F_c}$ be defined in the same way as in the proof of Lemma~\ref{functional_folding_use_almost_sat}. For $1 \leq i,j \leq \dim(\alpha)$ we can write
		\begin{equation*}
			\hat{g}(\alpha)_{ij}= \E_{{\bf x} \in G^{R}}\left[ g({\bf x}) \alpha({\bf x})_{ij} \right] = \frac{1}{|G|}\E_{{\bf x} \in G^{R}}\left[\sum_{c \in G} g(\vec{F_c} {\bf x}) \alpha(\vec{F_c}{\bf x})_{ij} \right]. 
		\end{equation*}
		We now fix ${\bf x} \in G^{R}$ 	and study the term $\sum_{c \in G} g(\vec{F_c} {\bf x}) \alpha(\vec{F_c}{\bf x})_{ij}$. Denote with ${\bf y^{c}}:=\vec{F_c}{\bf x}$, and observe that for each $c \in G$ we have  ${\bf y^{c}} \sim {\bf x}$  since
		\begin{equation*}
			y^{c}_d = F_c(\mathscr{p}_1(d),\hdots,\mathscr{p}_r(d)) x_d, \quad (\forall d \in [R]).
		\end{equation*}
			Hence, since $f$ is functionally folded we have that
			\begin{equation*}
				g(\vec{F_c} {\bf x} ) = \rho(f(\vec{F_c}{\bf x})_{pq} =\rho(c(f({\bf x}))_{pq}.
			\end{equation*}
			Let us now study the term $\alpha(\vec{F_c}{\bf x})_{ij}$. By Lemma~\ref{tensor_repr_G_n} we can write $\alpha({\bf x})$ as $\alpha({\bf x})=\alpha_1(x_1) \otimes \alpha_2(x_2) \otimes \hdots\otimes \alpha_R(x_R)$. For each $d \in [R]$ such that $\dim(\alpha_d) \geq 2$ by the assumption of the lemma we have that $(\mathscr{p}_1(d),\hdots,\mathscr{p}_r(d)) \neq {\bf b}$. Hence, for such $d$  we have that $F_c(\mathscr{p}_1(d),\hdots,\mathscr{p}_r(d))=1_G$ and 
			\begin{equation*}
				y_d^{c} = F_c(\mathscr{p}_1(d),\hdots,\mathscr{p}_r(d)) x_d = x_d.
			\end{equation*}
			 Therefore, in the expression 
			\begin{equation*}
				\alpha(\vec{F_c}{\bf x})=\alpha_1(\vec{F_c}_1x_1) \otimes \alpha_2(\vec{F_c}_2x_2) \otimes \hdots\otimes \alpha_R(\vec{F_c}_Rx_R), 
			\end{equation*}
			each representation $\alpha_d$ of dimension greater than $1$ satisfies $\alpha_d(\vec{F_c}_dx_d)=\alpha_d(x_d)$, and in particular for each such $d$ the value of $\alpha_d(\vec{F_c}_dx_d)$ does not change with with $c$. Hence, for a fixed ${\bf x}$ if we denote with $J$ the set of all the indices $d \in [R]$ such that $(\mathscr{p}_1(d),\hdots,\mathscr{p}_r(d))={\bf b}$ we can write 
			\begin{equation*}
				\alpha(\vec{F_c}{\bf x})_{ij} = \prod_{d \in J} [\alpha_d(cx_d)]_{d_id_j} \cdot C_{\alpha,{\bf x}},
			\end{equation*}
		 where $C_{\alpha,{\bf x}} \in \mathbb{C}$ is a constant that depends on $\alpha,{\bf x},$ but does not depend on $c$, and $1\leq d_i,d_j \leq \dim(\alpha_d)$. Observe that for each $\alpha_d$, where $d \in J$, is one-dimensional. Hence, we can write 
		 \begin{equation*}
			\prod_{d \in J} [\alpha_d(cx_d)]_{d_id_j} = \prod_{d \in J} \alpha_d (cx_d)=  \prod_{d \in J} \alpha_d(c)\prod_{d \in J} \alpha_d (x_d).
		 \end{equation*}
			Finally, since the product of one-dimensional representations is a one-dimensional representation, we can define with $\beta(c)$ the representation on $G$ given by
			\begin{equation*}
				\beta(c)= 	\prod_{d \in J} \alpha_d(c).
			\end{equation*}
			Then, we have that $\tilde{\beta}\colon G^{R} \to \mathbb{C}$ defined by $\tilde{\beta}({\bf x}) = \overline{\beta(\bf x)}$ is also a one-dimensional representation as shown in Claim 2.25 in~\cite{DBLP:journals/eccc/BhangaleK20}. But then
		\begin{equation*}
			\begin{split}
				\sum_{c \in G} g(\vec{F_c} {\bf x}) \alpha(\vec{F_c}{\bf x})_{ij} = \sum_{c \in G} \rho(c f({\bf x}))_{pq} \beta(c) \prod_{d \in J} \alpha_d(x_d) \cdot C_{\alpha,{\bf x}}\\ 
				= \sum_{c \in G} \sum_{1 \leq r \leq \dim(\rho)} \rho(c)_{pr} \rho(f({\bf x}))_{rq} \overline{\tilde{\beta}(c)} \prod_{d \in J} \alpha_d(x_d)\cdot C_{\alpha,{\bf x}} \\
				=\sum_{1 \leq r \leq \dim(\rho)}\left(\sum_{c \in G}  \rho(c)_{pr}  \overline{\tilde{\beta}(c)}\right) \rho(f({\bf x}))_{rq}\prod_{d \in J} \alpha_d(x_d) \cdot C_{\alpha,{\bf x}}=0,
			\end{split}
		\end{equation*}
			where the last inequality holds because $\sum_{c \in G} \rho(c)_{pr} \overline{\tilde{\beta}(c)}= |G| \langle \rho_{pr}, \tilde{\beta} \rangle_{L^{2}(G)}=0$ by Lemma~\ref{orthogonality} and since $\tilde{\beta}$ and $\rho$ are of different dimensions, and hence non-isomorphic. 
	\end{proof}
\end{lemma}

\section{Hardness of Max-$3$-lin with Almost Perfect Completeness with Universal Factor Graphs}
\label{section:imperfect_3lin}
In this section we describe a reduction from $(r,t)$-smooth UFG Label Cover to instances of Max-3-Lin over a non-abelian $G$ with factor graph independent of ${\bf b}=(b_1,\hdots,b_m)$. Let us introduce some notation first. For each $w \in W$ let $f_w^{F} \colon G^{R}\to G$ be functionally folded function with respect to $(\left \{\mathscr{p}_i\right\}_{i=1}^{r},(b_{I_1(w)},\hdots,b_{I_r(w)}))$, and for each $v \in V$ let $f_v^{F}\colon G^{L} \to G$ be ``classically folded'', i.e.,~function $f_v^{F}$ satisfies 
\begin{equation*}
	f_v^{F}({\bf x}c) = f_v({\bf x}) c.
\end{equation*}
Finally, let $f_w\colon G^{R}\to G$ be a function without any restriction. When querying the function $f_w^{F}({\bf x})$ in the procedure we describe below, we actually do not query $f_w^{F}({\bf x})$ directly. Instead, we find the equivalence class of ${\bf x}$  and the fixed representative $\bar{\bf x}$ of the equivalence class $[{\bf x}]$. Since ${\bf x} \sim \bar{\bf x}$  we have that ${\bf x}_d = F(\mathscr{p}_1(d),\hdots,\mathscr{p}_r(d)) \bar{\bf x}_d$. Because $f_w^{F}$  is functionally folded we have 
\begin{equation*}
	f_w^{F}({\bf x}) = F(b_{I_w(1)},\hdots,b_{I_w(r)}) f_w^{F}(\bar{\bf x}),
\end{equation*}
and instead of $f_w^{F}({\bf x})$ we actually query $f_w^{F}(\bar{\bf x})$ from the prover $P_1$ and whenever we use $f_w^{F}({\bf x})$ we actually mean  $F(b_{I_w(1)},\hdots,b_{I_w(r)}) f_w^{F}(\bar{\bf x})$. Note that $f_w^{F}(\bar{\bf x})$ does not depend on the values of $b_{I_w(1)},\hdots,b_{I_w(r)}$. Similarly, when querying $f_v^{F}({\bf x}\cdot c)$ in the test below we use $f_v^{F}({\bf x}\cdot c)$ to mean $f_v^F({\bf \bar{x}}) c$. With this in mind let us now we describe the Max-3-Lin instance we reduce to as a distribution of constraints generated by the following algorithm.
\begin{itemize} 
	\item Sample an edge $e=(v,w)$ from $E$.
	\item Sample ${\bf y}$ uniformly at random from $G^{L}$.
	\item Sample ${\bf x}$ uniformly at random from  $G^{R}$.
	\item Set ${\bf z} \in G^{R}$ such that $z_i = x_i^{-1} \cdot \eta_i \cdot (y \circ \pi_e)_i^{-1} $, where $\eta_i=1$  w.p. $1-\varepsilon$, and $\eta_i \sim G$ w.p. $\varepsilon$.
	\item Test $f_w^{F}({\bf x}) f_w({\bf z}) f_v^{F}({\bf y}) = 1_G$.
\end{itemize}
Since $f_v^{F}$ and $f_w^{F}$ are folded, the test $f_w^{F}(z) f_w(x) f_v^{F}(y) = 1_G$  is actually
\begin{equation*}
	F(b_{I_w(1)},\hdots,b_{I_w(r)}) f_w^{F}(\bar{\bf x}) f_w({\bf z}) f_v^{F}(\bar{ \bf y})c = 1_G,
\end{equation*}
which by rearranging gives us 
\begin{equation*}
	f_w^{F}(\bar{\bf x}) f_w({\bf z}) f_v^{F}(\bar{ \bf y}) = F(b_{I_w(1)},\hdots,b_{I_w(r)})^{-1} c^{-1}.
\end{equation*}
Since $f_w^{F}(\bar{\bf x}),f_w({\bf z}),f_v^{F}(\bar{\bf y})$ do not depend on the values of $b_{I_w(1)},\hdots,b_{I_w(r)}$ we get a Max-3-Lin instance whose factor graph does not depend on $b_1,\hdots,b_m$, and hence it remains to prove the following theorem.
\begin{theorem}\label{main_theorem_almost_perfect_completeness}
	Let $\varepsilon, \delta>0$, let $\zeta = \frac{\log(2|G|^{3}/\delta)}{\varepsilon}$, and consider $(r,t)$-smooth UFG Label Cover $\Lambda_{UFG}$ obtained by the parallel repetition of $(1,0.51)$-UFG-NP-hard Max-TSA, where $t$ is chosen such that $\Lambda_{UFG}$ has soundness at most $\frac{\delta^{4}}{32 |G|^{12}}\zeta^{-1}$, and $r$ is chosen such that
	\begin{equation*}
		\frac{t}{r} \leq \frac{\delta^{2}}{4|G|^{6}\zeta^{2}}.
	\end{equation*}
	 If $I$ is the instance of Max-3-Lin produced by the procedure described above with $\Lambda_{UFG}$ as the starting point, then the following holds:
	\begin{itemize}
		\item \emph{Completeness:} If $\Opt(\Lambda_{UFG}) = 1$,	then $\Opt(I) \geq 1-\varepsilon$.
		\item \emph{Soundness} If $\Opt(\Lambda_{UFG}) \leq  \frac{\delta^{4}}{32 |G|^{12}}\zeta^{-1}$, then $\Opt(I) \leq 1/|G| + \delta.$ 
	\end{itemize}
\end{theorem}
We first show completeness. 
\begin{lemma}
	If $\Opt(\Lambda_{UFG}) = 1$ 	then $\Opt(I) \geq 1-\varepsilon$.
	\begin{proof}
		If $\Opt(\Lambda_{UFG})$ then there is an assignment $\sigma$ which satisfies all the constraints of $\Lambda_{UFG}$. Let $f_w^{F}(x_1,\hdots,x_n)$  be defined as 
		\begin{equation*}
			f_w^{F}(x_1,\hdots,x_R) = x_{\sigma(w)}.
		\end{equation*}
		Note that $f_w^{F}$ is functionally folded. In particular, if $(x_1,\hdots,x_R):={\bf x} \sim \bar{\bf x} = (\overline{x_1},\hdots,\overline{x_R})$, then by the definition there is a function $F\colon \left \{0,1\right\}^{r}\to G$  such that 
		\begin{equation*}
			x_d = F(\mathscr{p}_1(d),\hdots,\mathscr{p}_r(d)) \overline{x_d}, \quad \textrm{for every } d \in [R].
		\end{equation*}
		But then we have that 
		\begin{equation*}
			f_w^F({\bf x}) = 	x_{\sigma(w)} = F(\mathscr{p}_1(\sigma(w)),\mathscr{p}_2(\sigma(w)),\hdots,\mathscr{p}_r(\sigma(w))) \overline{x_{\sigma(w)}}.
		\end{equation*}
		Since $\sigma$ is a satisfying assignment we have that $\mathscr{p}_i(\sigma(w)) = b_{I_w(i)}$ for every $i \in [r]$, and therefore
		\begin{equation*}
			f_w^{F}({\bf x}) = 	F(b_{I_w(1)},\hdots,b_{I_w(r)}) \overline{x_{\sigma(w)}} = F(b_{I_w(1)},\hdots,b_{I_w(r)})f_w^{F}(\bar{\bf x}),
		\end{equation*}
		which is sufficient to conclude that $f_w^F$ is folded. \par
		We also define $f_w$ by $f_w(z_1,\hdots,z_R) = z_{\sigma(w)}$ and $f_v^{F}(y_1,\hdots,y_L) = y_{\sigma(v)}$. It is straightforward to check that $f_v^{F}$ is ``classically folded''. Hence, it remains to show that the probability that the test 
		\begin{equation*}
			f_w^{F}({\bf x}) f_w({\bf z}) f_v^{F}({\bf y}) = 1_G
		\end{equation*}
		is passed is at least $1-\varepsilon$. But in case $\eta_{\sigma(w)}=1$ which happens with probability at least $1-\varepsilon$ we have that $z_{\sigma(w)}=  x_{\sigma(w)}^{-1} y_{\sigma(u)}^{-1}$ and hence
		\begin{equation*}
			f_w^{F}({\bf x}) f_w({\bf z}) f_v^{F}({\bf y}) =   x_{\sigma(w)} \cdot x_{\sigma(w)}^{-1} y_{\sigma(u)}^{-1} \cdot y_{\sigma(u)} = 1_G,
		\end{equation*}
		and therefore the test passes with probability at least $1-\varepsilon$.
	\end{proof}
\end{lemma}
Let us now show the contrapositive of soundness from Theorem~\ref{main_theorem_almost_perfect_completeness}.
\begin{lemma}\label{soundness_almost_perfect_lemma}
	If $\Opt(I) > \frac{1}{|G|}+\delta$ then $\Opt(\Lambda_{UFG}) > \frac{\delta^{4}}{32 |G|^{12}}\zeta^{-1}$.
	\begin{proof}
		By Lemma~\ref{dim_rho_lemma} the test is passed with probability
		\begin{equation*}
			\begin{split}
				\E\limits_{\substack{e \in E \\ e =(v,w)}} \E_{{\bf x,y,z}} \left[\frac{1}{|G|} \sum_{\rho \in \Irrep(G)} \dim(\rho)\chi_{\rho}(f_w^{F}({\bf x})f_w({\bf z}) f_v^{F}({\bf y}))\right]  \\
				= \frac{1}{|G|}+ \frac{1}{|G|}\sum_{\substack{\rho\in \Irrep(G)\\ \rho \not \cong 1}} \E\limits_{\substack{e \in E \\ e =(v,w)}}  \E_{{\bf x,y,z}} \left[\dim(\rho) \chi_\rho(f_w^{F}({\bf x}) f_w({\bf z}) f_v^{F}({\bf y}))\right] \\
				\leq \frac{1}{|G|}+ \frac{1}{|G|}\sum_{\substack{\rho\in \Irrep(G)\\ \rho \not \cong 1}} \dim(\rho)\left| \E\limits_{\substack{e \in E \\ e =(v,w)}}  \E_{{\bf x,y,z}} \left[ \chi_{\rho}(f_w^{F}({\bf x}) f_w({\bf z}) f_v^{F}({\bf y}))\right]\right|.
			\end{split}
		\end{equation*}
		Since the probability of acceptance is at least $1/|G|+\delta$ and $\sum_{\rho \in \Irrep(G)} \dim(\rho)^{2}=|G|$, there is at least one $\rho \in \Irrep(G) \setminus\left \{1\right\}$ such that 
		\begin{equation*}
			\left| \E\limits_{\substack{e \in E \\ e =(v,w)}} \E_{{\bf x,y,z}} \left[ \chi_{\rho}(f_w^{F}({\bf x}) f_w({\bf z}) f_v^{F}({\bf y}))\right]\right| \geq  \frac{\delta}{|G|}.
		\end{equation*}
		Let us fix one such $\rho$, and expand $\E_{{\bf x,y,z}} \left[ \chi_{\rho}(f_w^{F}({\bf x}) f_w({\bf z}) f_v^{F}({\bf y}))\right]$ as follows	
		\begin{equation*}
			\begin{split}
				\E_{{\bf x,y,z}} \left[\chi_{\rho}(f_w^{F}({\bf x}) f_w({\bf z}) f_v^{F}({\bf y}))\right] 
				= \E_{{\bf x,y,z}} \left[ \tr\left(\rho(f_w^{F}({\bf x}) f_w({\bf z}) f_v^{F}({\bf y}))\right)\right] \\
				= \E_{{\bf x,y,z}} \left[ \tr\left(\rho(f_w^{F}({\bf x}))\rho( f_w({\bf z})) \rho( f_v^{F}({\bf y}))\right)\right]  
				= \sum_{1\leq p,q,r \leq \dim(\rho)} \E \left[\rho(f_w^{F}({\bf x}))_{pq} \rho(f_w({\bf z}))_{qr} \rho(f_v^{F}({\bf y}))_{rp}\right].
			\end{split}
		\end{equation*}
		Hence, for at least one triple $p,q,r$ we have that 	
		\begin{equation}\label{reduced_form}
			\left| \E\limits_{\substack{e \in E \\ e =(v,w)}}  \E_{{\bf x,y,z}} \left[\rho(f_w^{F}({\bf x}))_{pq} \rho(f_w({\bf z}))_{qr} \rho(f_v^{F}({\bf y}))_{rp}\right] \right| \geq \delta / |G|^{3}.
		\end{equation}
		Let us now fix such $p,q,r$, and for the sake of notational convenience define 
		\begin{equation*}
			\begin{split}
				g_F({\bf x}) &= \rho(f_w^{F}({\bf x}))_{pq}, \\
				g({\bf z}) &= \rho(f_w({\bf z}))_{qr},\\
				h_F({\bf y}) &= \rho(f_v^{F}({\bf y}))_{rp}.
			\end{split}	
		\end{equation*}
		We can then write (\ref{reduced_form}) as
		\begin{equation} \label{soundness_among_neighbours}
			\left| \E\limits_{\substack{e \in E \\ e =(v,w)}}  \E_{{\bf x,y,z}} \left[ g_F({\bf x})g({\bf z}) h_F({\bf y})\right] \right| \geq \delta / |G|^{3}.
		\end{equation}
		We now study the term $\Theta^e := \E_{{\bf x,y,z}} \left[ g_F({\bf z})g({\bf x}) h_F({\bf y})\right]$. Note that the functions $h_F,g,g_F$ depend on the edge $e=(v,w)$ as well, but in the next couple of paragraphs since we consider $e$ to be fixed we will omit this dependence in our notation in order to avoid notational clutter. If we denote with ${\bf \eta}$ the random variable ${\bf \eta}=(\eta_1,\hdots,\eta_R)$, where $\eta_d$ are random variables defined in the beginning of this section, then by using the definition of ${\bf z}$ we have that 
		\begin{equation*}
			\begin{split}
				\Theta^e = \E_{{\bf x,y,z}} \left[ g_F({\bf x})g({\bf z}) h_F({\bf y})\right] = \E_{\bf x,y, \eta} \left[g_F({\bf x}) g({\bf x}^{-1}\cdot {\bf \eta} \cdot ({\bf y} \circ \pi_e)^{-1} )h_F({\bf y})\right]  \\
				= \E_{\bf y,\eta} \left[  (g_F*g) (\eta	\cdot  ({\bf y}^{-1}\circ \pi_e)  ) h_F({\bf y})\right]. 
			\end{split}
		\end{equation*}
		Now, by taking the Fourier transform of $h_F,g,$ and $g_F$, we can write the previous expression as
		\begin{equation*}
			\begin{split}
				\Theta^e = \E_{\bf y, \eta}\left[\left(\sum_{\alpha \in \Irrep(G^{R}) } \dim(\alpha) \left \langle \widehat{g_F} ( \alpha) \widehat{g}(\alpha), \alpha(\eta \cdot ({\bf y}^{-1}\circ \pi_e)   \right \rangle \right)  
				\left(\sum_{\beta \in \Irrep({G}^{L})} \dim(\beta) \langle \widehat{h_F}(\beta),\beta({\bf y}) \rangle \right) \right] \\
				= \E_{{\bf y, \eta}}\left[\left(\sum_{\alpha \in \Irrep(G^{R}) } \dim(\alpha) \left \langle \widehat{g_F} ( \alpha) \widehat{g}(\alpha), \alpha(\eta) \cdot \alpha({\bf y}^{-1}\circ \pi_e)   \right \rangle \right)  
				\left(\sum_{\beta \in \Irrep({G}^{L})} \dim(\beta) \langle \widehat{h_F}(\beta),\beta({\bf y}) \rangle \right) \right]. 
			\end{split}
		\end{equation*}
		Because $\E_{\eta}[\alpha(\eta)] = (1-\varepsilon)^{|\alpha|} \cdot Id$, we can rewrite the expression above as
		\begin{equation*} 
			\Theta^e= \E_{{\bf y}}\left[\left(\sum_{\alpha \in \Irrep(G^{R}) } \dim(\alpha) \left \langle \widehat{g_F} ( \alpha) \widehat{g}(\alpha), \alpha({\bf y}^{-1}\circ \pi_e)   \right \rangle (1-\varepsilon)^{|\alpha|} \right)  
				\left(\sum_{\beta \in \Irrep({G}^{L}) } \dim(\beta) \langle \widehat{h_F}(\beta),\beta({\bf y}) \rangle \right) \right].
		\end{equation*}
		Now, we expand the expressions $\left \langle \widehat{g_F} ( \alpha) \widehat{g}(\alpha), \alpha({\bf y}^{-1}\circ \pi_e)   \right \rangle$ and $\langle \widehat{h_F}(\beta),\beta({\bf y}) \rangle$ and get 
		\begin{equation} \label{soundness_before_study}
			\begin{split}
				\Theta^e=\E_{\bf y}\left[ \sum_{\substack{\alpha \in \Irrep(G^{R}) \\ \beta \in \Irrep(G^{L})}} \dim(\alpha)\dim(\beta) \sum_{ \substack{1 \leq i,j,k \leq \dim(\alpha)\\ 1 \leq i',j' \leq \dim(\beta)}} \widehat{g_F}(\alpha)_{ij} \widehat{g}(\alpha)_{jk} \cdot \alpha({\bf y} \circ \pi_e)_{ki} (1-\varepsilon)^{|\alpha|} \widehat{h_F}(\beta)_{i'j'} \beta({\bf y}^{-1})_{j'i'} \right]  \\
				=  \sum_{\substack{\alpha \in \Irrep(G^{R}) \\ \beta \in \Irrep(G^{L})}} \dim(\alpha)\dim(\beta) \sum_{ \substack{1 \leq i,j,k \leq \dim(\alpha)\\ 1 \leq i',j' \leq \dim(\beta)}} \widehat{g_F}(\alpha)_{ij} \widehat{g}(\alpha	)_{jk}  (1-\varepsilon)^{|\alpha|} \widehat{h_F}(\beta)_{i'j'} \E_{\bf y} \left[ \alpha({\bf y} \circ \pi_e)_{ki} \beta({\bf y}^{-1})_{j'i'} \right].
			\end{split}
		\end{equation}
		Consider the function $\alpha\circ \tilde{\pi}_e({\bf y}) = \alpha({\bf y} \circ \pi_e)$. As discussed in Section~\ref{subsection:reptheory}, homomorphism $\alpha \circ \tilde{\pi}_e ({\bf y})$ is (not necessarily irreducible) representation of $G^{L}$, and therefore $\alpha\circ \tilde{\pi}_e({\bf y}) \cong \oplus_{s=1}^{u} \tau^{s}({\bf y})$, where $\tau^{s}$  are irreducible representations of $G^{L}$. Hence, if $(k,i)$-th matrix entry of $\alpha\circ \tilde{\pi}_e$ does not belong to any block diagonal matrix $\tau^{s}$ we have that $\alpha({\bf y} \circ \pi_e)_{ki}=0$. Otherwise, we can let $(\beta',d_1,d_2) = Q(\alpha,k,i)$ and calculate
		\begin{equation*}
			\E_{\bf y} \left[ \alpha({\bf y} \circ \pi_e)_{ki} \beta({\bf y}^{-1})_{j'i'} \right] = \E_{\bf y} [ \beta'({\bf y})_{d_1d_2} \beta({\bf y}^{-1})_{j'i'}] = \E_{\bf y} [ \beta'({\bf y})_{d_1d_2} \overline{\beta({\bf y})_{i'j'}}],
		\end{equation*}
		where in the last equality we used the fact that $\beta$  is unitary representation. Now, by Theorem~\ref{orthogonality} we have that 
		\begin{equation*}
			\E_{\bf y} [ \beta'({\bf y})_{d_1d_2} \overline{\beta({\bf y})_{i'j'}}] \neq 0,
		\end{equation*}
		only if $Q(\alpha,k,i) = (\beta,i',j')$. In case $Q(\alpha,k,i)=(\beta,i',j')$, again by Theorem~\ref{orthogonality}, we have that
		\begin{equation*}
			\E_{\bf y} \left[ \alpha({\bf y} \circ \pi_e)_{ki} \beta({\bf y}^{-1})_{j'i'} \right] = 			\E_{\bf y} [ \beta'({\bf y})_{d_1d_2} \overline{\beta({\bf y})_{i'j'}}] = \frac{1}{\dim(\beta)}.
		\end{equation*}
		Hence, going back to (\ref{soundness_before_study}) we can write 
		\begin{equation*}
			\begin{split}	
				\Theta^e = \sum_{\substack{\alpha \in \Irrep(G^{R}) \\ \beta \in  \Irrep(G^{L}) }} \dim(\alpha)\dim(\beta) \sum_{ \substack{1 \leq i,j,k \leq \dim(\alpha)\\ 1 \leq i',j' \leq \dim(\beta)}} \widehat{g_F}(\alpha)_{ij} \widehat{g}(\alpha)_{jk}  (1-\varepsilon)^{|\alpha|} \widehat{h_F}(\beta)_{i'j'} \E_{\bf y} \left[ \alpha({\bf y} \circ \pi_e)_{ki} \beta({\bf y}^{-1})_{j'i'} \right] \\
				=\sum_{\alpha \in \Irrep(G^{R})  }  \sum_{1 \leq j,k \leq \dim(\alpha)} \dim(\alpha)\widehat{g}(\alpha)_{jk} \sum_{ \substack{i \in \B(\alpha,k)\\ (\beta,i',j')=Q(\alpha,k,i)}}  \widehat{g_F}(\alpha)_{ij}  (1-\varepsilon)^{|\alpha|} \widehat{h_F}(\beta)_{i'j'} .
			\end{split}	
		\end{equation*}
		Applying Cauchy-Schwarz inequality gives us 
		\begin{equation*}
			\begin{split}
				|\Theta^e|^{2} \leq \left(\sum_{\alpha \in \Irrep(G^{R})  } \dim(\alpha) \sum_{1 \leq j,k \leq \dim(\alpha)} |\hat{g}(\alpha)_{jk}|^{2}\right) \cdot \\
				\left(\sum_{\alpha \in \Irrep(G^{R})  } \sum_{1 \leq i,j\leq \dim(\alpha)} \dim(\alpha) \left|\sum_{ \substack{i \in \B(\alpha,k)\\ (\beta,i',j')=Q(\alpha,k,i)}}  \widehat{g_F}(\alpha)_{ij}  (1-\varepsilon)^{|\alpha|} \widehat{h_F}(\beta)_{i'j'} \right|^{2}\right).
			\end{split}
		\end{equation*}
		Now by Parseval's identity we have that 
		\begin{equation*}
			\sum_{\alpha \in \Irrep(G^{R})  } \dim(\alpha) \sum_{1 \leq j,k \leq \dim(\alpha)} |\hat{g}(\alpha)_{jk}|^{2} = \|g\|_{L_2(G^{R})}^{2} \leq 1,
		\end{equation*}
		where the last equality holds since $\rho$ is unitary. Therefore, we can write 	
		\begin{equation*}
			|\Theta^e|^{2}  \leq \sum_{\alpha \in \Irrep(G^{R})  } \sum_{1 \leq j,k \leq \dim(\alpha)} \dim(\alpha) \left|\sum_{ \substack{i \in \B(\alpha,k)\\ (\beta,i',j')=Q(\alpha,k,i)}}  \widehat{g_F}(\alpha) _{ij} (1-\varepsilon)^{|\alpha|} \widehat{h_F}(\beta)_{i'j'} \right|^{2}.
		\end{equation*}
		By applying Cauchy-Schwarz inequality to the inner term we obtain
		\begin{equation*}
			\begin{split}
				|\Theta^e|^{2} \leq \sum_{\alpha \in \Irrep(G^{R})  } \sum_{1 \leq j,k \leq \dim(\alpha)} \dim(\alpha)  
				\left(\sum_{ \substack{i \in \B(\alpha,k)\\ (\beta,i',j')=Q(\alpha,k,i)}} |\widehat{g_F}(\alpha)_{ij}|^{2}  (1-\varepsilon)^{2|\alpha|} \right)
				\left(\sum_{ \substack{ l \in \B(\alpha,k)\\ (\beta,i',j')=Q(\alpha,k,l)}} \left|\widehat{h_F}(\beta)_{i'j'}\right|^{2}\right) \\
				= \sum_{\alpha \in \Irrep(G^{R})  } \sum_{1 \leq i,j \leq \dim(\alpha)} \dim(\alpha)  |\widehat{g_F}(\alpha)_{ij}|^{2} (1-\varepsilon)^{2|\alpha|} 
					\left(\sum_{ \substack{ l,k \in \B(\alpha,i) \\ (\beta,i',j')=Q(\alpha,k,l)}} \left|\widehat{h_F}(\beta)_{i'j'}\right|^{2}\right).
			\end{split}
		\end{equation*}
		By Lemma~\ref{relationship_between_fourier_coeff_imperfect_completeness} we can upper bound this expression as 
		\begin{equation} \label{bound_before_case_analysis}
			\begin{split}
				|\Theta^e|^{2}\leq \sum_{\alpha \in \Irrep(G^{R})  } \sum_{1 \leq i,j \leq \dim(\alpha)} \dim(\alpha) |\widehat{g_F}(\alpha)_{ij}|^{2}(1-\varepsilon)^{2|\alpha|}  
				\sum_{\substack{ \beta \in \Irrep(G^{L})\\ T_{\pi_e}^{uq}(\alpha) \subseteq  T(\beta) \subseteq T_{\pi_e}(\alpha)	}   }   \|\widehat{h_F}(\beta)\|_{HS}^{2}  \\
					 \leq \sum_{\alpha \in \Irrep(G^{R})  } \sum_{\substack{ \beta \in \Irrep(G^{L})\\ T_{\pi_e}^{uq}(\alpha) \subseteq  T(\beta) \subseteq T_{\pi_e}(\alpha)	}} (1-\varepsilon)^{2|\alpha|}\dim(\alpha) \|\widehat{g_F}(\alpha)\|_{HS}^{2} \dim(\beta)				 \|\widehat{h_F}(\beta)\|_{HS}^{2}.
			\end{split}
		\end{equation}
		Recall that from (\ref{soundness_among_neighbours}) we have that 
		\begin{equation*}
				|\E\limits_{e \in E}	\Theta^e | \geq \delta/|G|^{3}.
		\end{equation*}
		By convexity we have that $	\E_{e	\in E}|\Theta^e |^{2} \geq |\E_e	\Theta^e |^{2}$, and therefore
		\begin{equation*}
			\E\limits_{e\in E}	|\Theta^e |^{2} \geq \delta^{2}/|G|^{6},
		\end{equation*}
		which with the expression in (\ref{bound_before_case_analysis}) gives us 
		\begin{equation*}
			\frac{\delta^{2}}{|G|^{6}} \leq \E_{e \in E}\left[\sum_{\alpha \in \Irrep(G^{R})  } \sum_{\substack{ \beta \in \Irrep(G^{L})\\ T_{\pi_e}^{uq}(\alpha) \subseteq  T(\beta) \subseteq T_{\pi_e}(\alpha)	}} (1-\varepsilon)^{2|\alpha|}\dim(\alpha) \|\widehat{g_F}(\alpha)\|_{HS}^{2} \dim(\beta)				 \|\widehat{h_F}(\beta)\|_{HS}^{2}\right].
		\end{equation*}
		The contribution of the terms with $|\alpha|>\zeta$ where $\zeta=\frac{\log(2|G|^{3}/\delta)}{ \varepsilon}$ is small, since in that case $(1-\varepsilon)^{2|\alpha|} \leq \frac{\delta^{2}}{4|G|^{6}}$, and hence 
		\begin{equation*}
			\begin{split}
				\E_{e \in E}\left[\sum_{\substack{\alpha \in \Irrep(G^{R})\\|\alpha| \geq \zeta } } \sum_{\substack{ \beta \in \Irrep(G^{L})\\ T_{\pi_e}^{uq}(\alpha) \subseteq  T(\beta) \subseteq T_{\pi_e}(\alpha)	}} (1-\varepsilon)^{2|\alpha|}\dim(\alpha) \|\widehat{g_F}(\alpha)\|_{HS}^{2} \dim(\beta)				 \|\widehat{h_F}(\beta)\|_{HS}^{2}\right] \\
				\leq 		\frac{\delta^{2}}{4|G|^{6}}	\E_{e \in E}\left[\sum_{\substack{\alpha \in \Irrep(G^{R})\\|\alpha| \geq \zeta } } \sum_{\substack{ \beta \in \Irrep(G^{L})\\ T_{\pi_e}^{uq}(\alpha) \subseteq  T(\beta) \subseteq T_{\pi_e}(\alpha)	}} \dim(\alpha) \|\widehat{g_F}(\alpha)\|_{HS}^{2} \dim(\beta)				 \|\widehat{h_F}(\beta)\|_{HS}^{2}\right]  \\
				\leq \frac{\delta^{2}}{4|G|^{6}} \|g_F\|_{L^{2}(G^{R})}^{2} \|h_F\|_{L^{2}(G^{R})}^{2} \leq \frac{\delta^{2}}{4|G|^{6}}. 
			\end{split}
		\end{equation*}
		Therefore	we have 
		\begin{equation*}
			\frac{3\delta^{2}}{4|G|^{6}} \leq \E_{e \in E}\left[\sum_{\substack{\alpha \in \Irrep(G^{R})\\ |\alpha| <\zeta  }} \sum_{\substack{ \beta \in \Irrep(G^{L})\\ T_{\pi_e}^{uq}(\alpha) \subseteq  T(\beta) \subseteq T_{\pi_e}(\alpha)	}} \dim(\alpha) \|\widehat{g_F}(\alpha)\|_{HS}^{2} \dim(\beta)				 \|\widehat{h_F}(\beta)\|_{HS}^{2}\right].
		\end{equation*}
		As in Lemma~\ref{smoothness_lemma} let us use $E_w$ to denote the set of of all edges $e$  incident to $w$. If we denote the inner term with $t^e(\alpha,\beta) :=\dim(\alpha) \|\widehat{g_F}(\alpha)\|_{HS}^{2} \dim(\beta)				 \|\widehat{h_F}(\beta)\|_{HS}^{2}$, we can write 
		\begin{equation*}
			\begin{split}
				\frac{3\delta^{2}}{4|G|^{6}} \leq \E_{w \in  W} \E_{e \in E_w} \left[\sum_{\substack{\alpha \in \Irrep(G^{R})\\ |\alpha| <\zeta  }} \sum_{\substack{ \beta \in \Irrep(G^{L})\\ T_{\pi_e}^{uq}(\alpha) \subseteq  T(\beta) \subseteq T_{\pi_e}(\alpha)	}} t^e(\alpha,\beta) \right] = I_1 + I_2\\
			\end{split}
		\end{equation*}
		where 
		\begin{equation*}
			I_1 := 					\E_{w \in W}\E_{e \in E_w} \left[\sum_{\substack{\alpha \in \Irrep(G^{R})\\ |\alpha| <\zeta  }} \sum_{\substack{ \beta \in \Irrep(G^{L})\\ T_{\pi_e}^{uq}(\alpha) \subseteq  T(\beta) \subseteq T_{\pi_e}(\alpha)	}} t^e(\alpha,\beta) \mathcal{C}(T(\alpha),\pi_e) \right],
		\end{equation*}
		and
		\begin{equation*}
			I_2 := \E_{w \in W}\E_{e \in E_w}\left[\sum_{\substack{\alpha \in \Irrep(G^{R})\\ |\alpha| <\zeta  }} \sum_{\substack{ \beta \in \Irrep(G^{L})\\ T_{\pi_e}^{uq}(\alpha) \subseteq  T(\beta) \subseteq T_{\pi_e}(\alpha)	}} t^e(\alpha,\beta)(1-\mathcal{C}(T(\alpha),\pi_e)) \right] .
		\end{equation*}
		Let us now show that the contribution of the term $I_1$ is small due to the smoothness of $\Lambda_{UFG}$. By unwrapping the definition of $t^e(\alpha,\beta)$  we have 
		\begin{equation*}
			\begin{split}
				I_1 = \E_{w \in W}\E_{e \in E_W} \left[\sum_{\substack{\alpha \in \Irrep(G^{R})\\ |\alpha| <\zeta  }} \sum_{\substack{ \beta \in \Irrep(G^{L})\\ T_{\pi_e}^{uq}(\alpha) \subseteq  T(\beta) \subseteq T_{\pi_e}(\alpha)	}} t^e(\alpha,\beta) \mathcal{C}(T(\alpha),\pi_e) \right]   \\ 
				=		\E_{w \in W}\E_{e \in E_w} \left[\sum_{\substack{\alpha \in \Irrep(G^{R})\\ |\alpha| <\zeta  }} \mathcal{C}(T(\alpha),\pi_e) \dim(\alpha) \|\widehat{g_F}(\alpha)\|_{HS}^{2} \sum_{\substack{ \beta \in \Irrep(G^{L})\\ T_{\pi_e}^{uq}(\alpha) \subseteq  T(\beta) \subseteq T_{\pi_e}(\alpha)	}}  \dim(\beta)				 \|\widehat{h_F}(\beta)\|_{HS}^{2}   \right] \leq  \\
				\E_{w \in W}\E_{e \in E_w} \left[\sum_{\substack{\alpha \in \Irrep(G^{R})\\ |\alpha| <\zeta  }} \mathcal{C}(T(\alpha),\pi_e) \dim(\alpha) \|\widehat{g_F}(\alpha)\|_{HS}^{2} \|h\|_{L^{2}(G^{L})}  \right] \\
				\leq\E_{w \in W}\E_{e \in E_w} \left[\sum_{\substack{\alpha \in \Irrep(G^{R})\\ |\alpha| <\zeta  }} \mathcal{C}(T(\alpha),\pi_e) \dim(\alpha) \|\widehat{g_F}(\alpha)\|_{HS}^{2}  \right].
			\end{split}
		\end{equation*}
		But then by Lemma~\ref{smoothness_lemma} we have that
			\begin{equation*}
				\begin{split}
					I_1 \leq \E_{w \in W}\left[\sum_{\substack{\alpha \in \Irrep(G^{R})\\ |\alpha| <\zeta  }}\dim(\alpha) \|\widehat{g_F}(\alpha)\|_{HS}^{2} \E_{e \in E_w} \left[ \mathcal{C}(T(\alpha),\pi_e) \right] \right]  \\
					 \leq \E_{w \in W}\left[ \sum_{\substack{\alpha \in \Irrep(G^{R})\\ |\alpha| <\zeta  }} \dim(\alpha) \|\widehat{g_F}(\alpha)\|_{HS}^{2}|\alpha|^{2}\frac{t}{r}\right] \leq \zeta^{2}\frac{t}{r}.
				\end{split}
			\end{equation*}
			where in the last inequality we used the fact that $\sum_{\alpha \in \Irrep(G^{R})}\dim(\alpha) \|\widehat{g_F}(\alpha)\|_{HS}^{2} \leq \|g_F\|_{L^{2}(G^{R})}^{2} \leq 1 $. Finally, due to our choice of $r,t$  we have that
				\begin{equation*}
					I_1 \leq \frac{\delta^{2}}{4 |G|^{6}},
				\end{equation*}
				and hence 
				\begin{equation*}
					\frac{\delta^{2}}{2 |G|^{6}} \leq I_2.
				\end{equation*}
				Recalling our definition of $I_2$ we can summarize our work so far by writing
				\begin{equation*}
					\frac{\delta^{2}}{2 |G|^{6}} \leq \E_{e \in E}\left[\sum_{\substack{\alpha \in \Irrep(G^{R})\\ |\alpha| <\zeta, \mathcal{C}(T(\alpha),\pi_e)=0 }} \sum_{\substack{ \beta \in \Irrep(G^{L})\\ T_{\pi_e}^{uq}(\alpha) \subseteq  T(\beta) \subseteq T_{\pi_e}(\alpha)	}} t^e(\alpha,\beta)  \right].
				\end{equation*}
				Hence, by Markov's inequality there is $E' \subseteq E, |E'| \geq \frac{\delta^{2}}{4 |G|^{6}} |E| $ such that for each $e \in E'$ we have 
				\begin{equation}\label{final_bound}
					\frac{\delta^{2}}{4 |G|^{6}} \leq \sum_{\substack{\alpha \in \Irrep(G^{R})\\ |\alpha| <c, \mathcal{C}(T(\alpha),\pi_e)=0  }} \sum_{\substack{ \beta \in \Irrep(G^{L})\\ T_{\pi_e}^{uq}(\alpha) \subseteq  T(\beta) \subseteq T_{\pi_e}(\alpha)	}} \dim(\alpha) \|\widehat{g_F}(\alpha)\|_{HS}^{2} \dim(\beta)				 \|\widehat{h_F}(\beta)\|_{HS}^{2} .
				\end{equation}
		Consider now the following randomized labelling of the starting UFG Label Cover instance.	For a vertex $w$  we pick a label $\sigma_w$ as follows. First, we pick a representation $\alpha$ with probability $\dim(\alpha)\|\widehat{g_F}(\alpha)\|_{HS}^{2}$. Then, if we consider the decomposition $\alpha=(\alpha_1,\hdots,\alpha_R)$, we select $\sigma_w$ uniformly at random from the set 
	\begin{equation*}
		\left \{d \in [R] \mid \alpha_d \not \cong 1 \textrm{ and for every } i \in [r],  \mathscr{p}_i(d)=b_{I_w(i)}\right\}.
	\end{equation*}
	Observe that whenever $\widehat{g_F}(\alpha)\neq 0$ the set above is not empty by Lemma~\ref{functional_folding_use_almost_sat}, and therefore we can always make such a choice. For a vertex $v$  we pick a label $\sigma_v$ in analogous way, i.e.,~we first pick a representation $\beta$ with probability $\dim(\beta)\|\widehat{h_F}(\beta)\|_{HS}^{2}$. Then, if we consider the decomposition $\beta=(\beta,\hdots,\beta_L)$, we select $\sigma_v$ uniformly at random from the set $\left \{d \in [L] \mid \beta_d \not \cong 1\right\}$. Note again that this set is not empty by Lemma~\ref{trivial_fourier_0_folding_lemma}. Then, by inequality (\ref{final_bound}) for each $e \in E'$ we have that $\alpha,\beta$ satisfy with probability at least $\frac{\delta^{2}}{4 |G|^{6}}$ the following properties
	\begin{itemize}
		\item $|\alpha| <\zeta$,
		\item $\mathcal{C}(T(\alpha),\pi_e)=0$,
		\item $T_{\pi_e}^{uq}(\alpha) \subseteq  T(\beta) \subseteq T_{\pi_e}(\alpha)$.
	\end{itemize}
	Let us now condition on this event happening and calculate the probability that the agreement test is passed along the edge $e$. Observe that since $ T(\beta) \subseteq T_{\pi_e}(\alpha)$ we have that $|\beta| \leq |\alpha| \leq \zeta$. Furthermore, since $\mathcal{C}(T(\alpha),\pi_e)=0$ we have that $\pi_e(\sigma_w) \in T_{\pi_e}^{uq}(\alpha)$ and hence since $T(\beta) \supseteq T_{\pi_e }^{uq}(\alpha)$ the probability that the test is passed is at least 
	\begin{equation*}
		\frac{1}{|\beta|}	\geq \frac{1}{|\alpha|}  \geq \zeta^{-1}.
	\end{equation*}
	Hence, for $e \in E'$ 	the test passes with probability at least 
	\begin{equation*}
		\frac{\delta^{2}}{4 |G|^{6}} \zeta^{-1}, 
	\end{equation*}
	and since $|E'| \geq \frac{\delta^{2}}{4 |G|^{6}} |E|$ the randomized assignment in expectation satisfies at least 
	\begin{equation*}
		\frac{\delta^{4}}{16 |G|^{12}} \zeta^{-1}
	\end{equation*}
	fraction of edges. But the observe that by the method of conditional expectations there is a labelling that satisfies at least $\frac{\delta^{4}}{16 |G|^{4}}\zeta^{-1}$ fraction of edges, which is a contradiction with the assumption that $\Opt(\Lambda_{UFG}) \leq \frac{\delta^{4}}{32 |G|^{12}} \zeta^{-1}$. Hence, the initial assumption in our argument is incorrect, and therefore $\Opt(I) \leq 1/|G|+\delta$.
	\end{proof}
\end{lemma}
\section{Hardness of Max-$3$-Lin with Perfect Completeness with Universal Factor Graphs}
\label{section:perfect_3lin}
In this section we discuss the hardness result for Max-3-Lin with perfect completeness. Since many steps in this result share a lot of similarities with the proof from the previous section and with~\cite{DBLP:journals/eccc/BhangaleK20}, we will keep the discussion in this section much shorter and focus mostly on the new techniques specific to the reduction introduced in this section.
\par 
In order to apply the argument from~\cite{DBLP:journals/eccc/BhangaleK20} for high-dimensional terms we will show that we can suitably pick $(r,t)$ such that $(r,t)$-smooth UFG Label Cover $\Lambda_{UFG}$ satisfies the same \emph{smoothness} property as the starting Label Cover instance in~\cite{DBLP:journals/eccc/BhangaleK20}. In particular, we have the following lemma.
\begin{lemma}\label{smoothness_bhangale_khot}
	Consider $(r,t)$-smooth UFG Label Cover $\Lambda_{UFG}$ constructed by the parallel repetition of $(1,0.51)$-UFG-NP-hard Max-TSA instance. There is a constant $d_0 \in (0,1/3)$ such that, given any fixed $t$, for all sufficiently big $r$  we have that
	\begin{equation*}
		\Pr_{e \in E_w} [ |\pi_e(S)| <|S|^{d_0} ] \leq  \frac{1}{|S|^{d_0}}, \quad (\forall w \in W, S \subseteq  [R] ).
	\end{equation*}
\end{lemma}
We provide the proof of this lemma in the appendix. We remark that one can also prove this lemma by applying Lemma 6.9 from~\cite{DBLP:journals/jacm/Hastad01} to our setting, in which case the statement holds for any $r\geq t$. However, since in this work we use parallel repetition of Max-TSA instead of Max-$3$-Sat we opt to reprove this lemma for the sake of completeness, and use less involved argument at the expense of having possibly much larger $r$ which increases the size of starting $\Lambda_{UFG}$ instance polynomially and hence does not affect the overall argument of this section.
\par
As in the previous section we start from $(r,t)$-smooth UFG Label Cover, and along each edge we test functions $f_w^F,f_w \colon G^{R} \to G, f_v \colon G^{L} \to G$, where $f_w^F$  and $f_v^{F}$  are folded in the same way as previously. The reduction to Max-3-Lin instance is given as a distribution of constraints generated by the following algorithm.
\begin{itemize} 
	\item Sample an edge $e=(v,w)$ from $E$.
	\item Sample ${\bf y}$ uniformly at random from $G^{L}$.
	\item Sample ${\bf x}$ uniformly at random from  $G^{R}$.
	\item Set ${\bf z} \in G^{R}$ such that $z_i = x_i^{-1} \cdot (y \circ \pi_e)_i^{-1} $.
	\item Test $f_w^{F}({\bf x}) f_w({\bf z}) f_v^{F}({\bf y}) = 1_G$.
\end{itemize}
This is essentially the same test as in~\cite{DBLP:journals/eccc/BhangaleK20}; the only difference comes from the fact that we are folding two instead of three functions and that we are using functional folding on one of the functions. Let us now state the main theorem of this section.
\begin{theorem}\label{main_theorem_perfect_completeness}
	Let $\varepsilon, \delta>0$, and $C$ be a constant such that $C^{-d_0/2} \leq \frac{\delta^{2}}{12|G|^{6}}$. Consider $(r,t)$-smooth UFG Label Cover $\Lambda_{UFG}$ obtained by the parallel repetition of $(1,0.51)$-UFG-NP-hard Max-TSA, where $t$ is chosen such that $\Lambda_{UFG}$ has soundness at most $\frac{\delta^{4}}{257 |G|^{12+2C}}C^{-1}$, and $r$ is chosen sufficiently big so that Lemma~\ref{smoothness_bhangale_khot} holds and 
	\begin{equation*}
		\frac{t}{r} \leq \frac{\delta^{2}}{8|G|^{6+C}C^{2}}.
	\end{equation*}
 If $I$ is the instance of Max-3-Lin produced by the procedure described above with $\Lambda_{UFG}$ as the starting point, then the following holds:
	\begin{itemize}
		\item \emph{Completeness:} If $\Opt(\Lambda_{UFG}) = 1$ 	then $\Opt(I) = 1$.
		\item \emph{Soundness} If $\Opt(\Lambda_{UFG}) \leq 	 \frac{\delta^{4}}{257 |G|^{12+2C}}C^{-1}$ then $\Opt(I) \leq 1/|[G,G]| + \delta$.
	\end{itemize}
	\begin{proof}
		As in the proof from the previous section the completeness case follows by setting $f_w^F, f_w$, and $f_v^{F}$ to be the dictator functions encoding the satisfying assignment to $\Lambda_{UFG}$. 
		\par 
		For the soundness case we closely follow the proof given in~\cite{DBLP:journals/eccc/BhangaleK20}. In particular, the test introduced in the beginning of this section is passed with probability 
		\begin{equation*}
			\begin{split}
			\E\limits_{\substack{e \in E \\ e =(v,w)}}  \E_{{\bf x,y,z}} \left[\frac{1}{|G|} \sum_{\rho \in \Irrep(G)} \dim(\rho)\chi_{\rho}(f_w^{F}({\bf x})f_w({\bf z}) f_v^{F}({\bf y}))\right]  \\
				= \frac{1}{|G|}\sum_{\substack{\rho \in \Irrep(G)\\ \dim(\rho)=1}} \E\limits_{\substack{e \in E \\ e =(v,w)}} \E_{{\bf x,y,z}} \left[\dim(\rho) \chi_\rho(f_w^{F}({\bf x}) f_w({\bf z}) f_v^{F}({\bf y}))\right]\\
				+ \frac{1}{|G|}\sum_{\substack{\rho \in \Irrep(G)\\ \dim(\rho)\geq 2}} \E\limits_{\substack{e \in E \\ e =(v,w)}} \E_{{\bf x,y,z}} \left[\dim(\rho) \chi_\rho(f_w^{F}({\bf x}) f_w({\bf z}) f_v^{F}({\bf y}))\right]\\
				\leq \frac{1}{|G|}\sum_{\substack{\rho \in \Irrep(G)\\ \dim(\rho)=1}} 1 +\frac{1}{|G|}\sum_{\substack{\rho \in \Irrep(G)\\ \dim(\rho)\geq 2}} \left| \E\limits_{\substack{e \in E \\ e =(v,w)}} \E_{{\bf x,y,z}} \left[\dim(\rho) \chi_\rho(f_w^{F}({\bf x}) f_w({\bf z}) f_v^{F}({\bf y}))\right]\right|  \\
				= \frac{1}{|[G,G]|}+\frac{1}{|G|}\sum_{\substack{\rho \in \Irrep(G)\\ \dim(\rho)\geq 2}} \left|\E\limits_{\substack{e \in E \\ e =(v,w)}} \E_{{\bf x,y,z}} \left[\dim(\rho) \chi_\rho(f_w^{F}({\bf x}) f_w({\bf z}) f_v^{F}({\bf y}))\right]\right|.
			\end{split}
		\end{equation*}
	Let us argue by contradiction and assume that 	$\Opt(I) > 1/|[G,G]| + \delta$ which allows us to fix $\rho$ such that $\dim(\rho)\geq 2$ and 
		\begin{equation*}
			\left|\E\limits_{\substack{e \in E \\ e =(v,w)}} \E_{{\bf x,y,z}} \left[ \chi_\rho(f_w^{F}({\bf x}) f_w({\bf z}) f_v^{F}({\bf y})\right]\right| \geq \frac{\delta}{|G|}.
		\end{equation*}
		By replacing $\chi_{\rho}$ with $\tr \circ \rho$ and expanding the expressions in the same way as in the proof of the Theorem~\ref{main_theorem_almost_perfect_completeness} we conclude that for some $1 \leq p,q,r \leq \dim(\rho)$ we have that 
		\begin{equation}\label{reduced_form2}
			\left| \E\limits_{\substack{e \in E \\ e =(v,w)}} \E_{{\bf x,y,z}} \left[\rho(f_w^{F}({\bf x}))_{pq} \rho(f_w({\bf z}))_{qr} \rho(f_v^{F}({\bf y}))_{rp}\right] \right| \geq \delta / |G|^3.
		\end{equation}
		Let us define $g_F,g\colon G^{R}\to \mathbb{C},h_F\colon G^{L}\to \mathbb{C}$ and $\Theta^e$ as before, and observe that by writing these functions in their respective Fourier basis we arrive at the expression
		\begin{equation*}
			\Theta^e=  \sum_{\substack{\alpha \in \Irrep(G^{R}) \\ \beta \in \Irrep(G^{L})}} \dim(\alpha)\dim(\beta) \E_{\bf y}\left[ \langle \widehat{g_F}(\alpha) \hat{g}(\alpha),\alpha({\bf y}^{-1} \circ \pi)\rangle  \langle \widehat{h_F}(\beta)\beta({\bf y})\rangle\right].
		\end{equation*}
		By Lemma~\ref{trivial_fourier_0_folding_lemma_perfect_completeness} we have for $\alpha \in \Irrep(G^{R})$ with $\dim(\alpha)=1$ that $\widehat{g_F}(\alpha)=0$, and for $\beta \in \Irrep(G^{L})$ with $\dim(\beta)=1$ that $\widehat{h_F}(\beta)=0$. Hence, by following the notation of~\cite{DBLP:journals/eccc/BhangaleK20} let us denote with $\Term^e(\alpha,\beta)$ the inner term
		\begin{equation*}
			\Term^e(\alpha,\beta)	=  \dim(\alpha)\dim(\beta) \E_{\bf y}\left[ \langle \hat{g_F}(\alpha) \hat{g}(\alpha),\alpha({\bf y}^{-1} \circ \pi)\rangle  \langle \widehat{h_F}(\beta)\beta({\bf y})\rangle \right],
		\end{equation*}
		and write 	
		\begin{equation*}
			\E_{e \in E} \left[\Theta^e\right] = \E_e \left[\Theta^{e}(\textrm{low})\right] + \E_{e \in E}\left[\Theta^{e}(\textrm{high})\right],
		\end{equation*}
		where 
		\begin{equation*}
			\begin{split}
				\Theta^{e}(\textrm{low}) = \sum_{\substack{\alpha \in \Irrep(G^{R}) \\ \beta \in \Irrep(G^{L}) \\ \dim(\alpha), \dim(\beta)\geq 2 \\ |\tilde{T}(\alpha)| \leq C}} \Term^{e}(\alpha,\beta), \\
				\Theta^{e}(\textrm{high}) = \sum_{\substack{\alpha \in \Irrep(G^{R}) \\ \beta \in \Irrep(G^{L}) \\ \dim(\alpha), \dim(\beta)\geq 2 \\ |\tilde{T}(\alpha)| > C}} \Term^{e}(\alpha,\beta).
			\end{split}	
		\end{equation*}
		 In order to arrive at a contradiction to $|\E_{e \in E}[ \Theta^e	]|\geq  \delta / |G|^{3}$ we will show that $\left|\E_{ e \in E} \Theta^{e}(\textrm{low})\right|< \frac{\delta}{2|G|^{3}}$ and $\left|\E_{e\in E} \Theta^{e}(\textrm{high})\right|\leq \frac{\delta}{2|G|^{3}}$. In order to bound $\E_{e \in E} \Theta^{e}(\textrm{high})$ due to our choice of the constant $C$ we can just observe that Claim 4.7.~from~\cite{DBLP:journals/eccc/BhangaleK20} can be applied in our setting to yield
		\begin{equation*}
			\left|\E_{e \in E}\Theta^{e}(\textrm{high})\right| \leq \frac{\delta}{2|G|^{3}}.
		\end{equation*}
		In particular, since the starting UFG Label Cover instance is smooth one can just replace $g,g',h$  in the proof of Claim 4.7.~with $g_F,g,h_F$ and check the correctness of all the steps in the proof for these functions. 
		\par 
		For the low degree terms we can use argument identical to the one in Claim 4.6. of~\cite{DBLP:journals/eccc/BhangaleK20} to arrive at the expression 
		\begin{equation}\label{final_bound_perfect_completeness_low_degree}
			\left|\Theta^{e}(\textrm{low})\right|^{2} \leq |G|^{C} \sum_{\substack{\alpha \in \Irrep(G^{R}) \\ \beta \in \Irrep(G^{L}) \\ \dim(\alpha), \dim(\beta)\geq 2 \\ |\tilde{T}(\alpha)| \leq C\\ \tilde{T}_{\pi_e}^{uq}(\alpha)\subseteq \tilde{T}(\beta) \subseteq \tilde{T}_{\pi_e}(\alpha) }} \dim(\alpha) \|\widehat{g_F}(\alpha)\|_{HS}^{2} \dim(\beta)				 \|\widehat{h_F}(\beta)\|_{HS}^{2}
.
		\end{equation}
		The relationship between $\tilde{T}(\beta) \subseteq \tilde{T}_{\pi_e}(\alpha)$ already appeared in Claim 4.6 of~\cite{DBLP:journals/eccc/BhangaleK20} where it was written as ``$\beta_{\geq 2} \subseteq \pi_{\geq 2}(\alpha)$''. On the other side, the relationship $\tilde{T}_{\pi_e}^{uq}(\alpha)\subseteq \tilde{T}(\beta)$ is new. The argument as to why we have $\tilde{T}_{\pi_e}^{uq}(\alpha)\subseteq \tilde{T}(\beta)\subseteq \tilde{T}_{\pi_e}(\alpha)$ is essentially the same as the argument for having $T_{\pi_e}^{uq}(\alpha)\subseteq T(\beta)\subseteq T_{\pi_e}(\alpha)$ in the proof of Theorem~\ref{main_theorem_almost_perfect_completeness}. In particular, the relationship between $\alpha,\beta$  appears because once we study the inner term $ \E_{\bf y}\left[ \langle \widehat{g_F}(\alpha) \hat{g}(\alpha),\alpha({\bf y}^{-1} \circ \pi)\rangle  \langle \widehat{h_F}(\beta)\beta({\bf y})\rangle\right]$ of $\Term^{e}(\alpha,\beta)$ and write out the scalar products between matrices in terms of their entries we get
		\begin{equation*}
			\begin{split}
\E_{\bf y}\left[ \langle \widehat{g_F}(\alpha) \hat{g}(\alpha),\alpha({\bf y}^{-1} \circ \pi)\rangle  \langle \widehat{h_F}(\beta)\beta({\bf y})\rangle\right]\\
				= \sum_{\substack{1\leq i,j,k\leq \dim(\alpha)\\1 \leq i',j'\leq \dim(\beta)}} \widehat{g_F}(\alpha)_{ij} \widehat{g}(\alpha	)_{jk}  \widehat{h_F}(\beta)_{i'j'} \E_{\bf y} \left[ \alpha\circ \tilde{\pi}_e({\bf y})_{ki} \beta({\bf y}^{-1})_{j'i'} \right],
			\end{split}
		\end{equation*}
		and hence non-zero terms need to have that $\E_{\bf y}\left[ \alpha\circ\tilde{\pi}_e({\bf y})_{ki} \beta({\bf y}^{-1})_{j'i'} \right] \neq 0$. However, since $\alpha \circ \tilde{\pi}_e$ is a representation of $G^{L}$, we can represent $ \alpha\circ \tilde{\pi}_e$ as $ \alpha\circ \tilde{\pi}_e=\oplus_{s=1}^{u} \tau^{s}$ where $\tau^s \in \Irrep(G^{L})$. In case $(k,i)$-th matrix entry of $\alpha\circ \tilde{\pi}_e$ does not belong to any block $\tau^s$ we have $\alpha\circ \tilde{\pi}_e({\bf y})=0$. Otherwise,
		\begin{equation*}
			\alpha({\bf y} \circ \pi_e)_{ki} = \beta'({\bf y})_{d_1d_2},
		\end{equation*}
		where $(\beta',d_1,d_2)=Q(\alpha,k,i)$. But then by Lemma~\ref{orthogonality} we have that 
		\begin{equation*}
			\E_{\bf y} [ \beta'({\bf y})_{d_1d_2} \beta({\bf y}^{-1})_{j'i'}] = \E_{\bf y} [ \beta'({\bf y})_{d_1d_2} \overline{\beta({\bf y})_{i'j'}}]\neq 0,
		\end{equation*}
		only if $Q(\alpha,k,i) = (\beta,i',j')$. But this implies that $\beta\cong \tau^{s}$ for some $s=1,\hdots,t$, and therefore by Lemma~\ref{relationship_between_fourier_coeff_perfect_completeness} we have that $\tilde{T}_{\pi_e}^{uq}(\alpha)\subseteq \tilde{T}(\beta) \subseteq \tilde{T}_{\pi_e}(\alpha)$. The rest of the proof is completely analogous to the proof given in Lemma~\ref{soundness_almost_perfect_lemma}. In particular, we argue by contradiction and assume that $\left|\E_{e \in E} \Theta^{e}(\textrm{low})\right|\geq  \frac{\delta}{2|G|^{3}}$. Then, by convexity we have that $\E_{e \in E} \left|\Theta^{e}(\textrm{low})\right|^{2} \geq \left|\E_{e \in E} \Theta^{e}(\textrm{low})\right|^{2}$ and hence 
			\begin{equation*}
				\E_{e \in E} \left|\Theta^{e}(\textrm{low})\right|^{2} \geq  \frac{\delta^{2}}{4|G|^{6}},
			\end{equation*}
	  which with (\ref{final_bound_perfect_completeness_low_degree}) gives us
			\begin{equation*}
			\frac{\delta^{2}}{4|G|^{6}}	\leq  |G|^{C} \sum_{\substack{\alpha \in \Irrep(G^{R}) \\ \beta \in \Irrep(G^{L}) \\ \dim(\alpha), \dim(\beta)\geq 2 \\ |\tilde{T}(\alpha)| \leq C\\ \tilde{T}_{\pi_e}^{uq}(\alpha)\subseteq \tilde{T}(\beta) \subseteq \tilde{T}_{\pi_e}(\alpha) }} \dim(\alpha) \|\widehat{g_F}(\alpha)\|_{HS}^{2} \dim(\beta)				 \|\widehat{h_F}(\beta)\|_{HS}^{2}.
		\end{equation*}
		Now, by our choice of $(r,t)$, and according to Lemma~\ref{smoothness_lemma}, the smoothness property of $\Lambda_{UFG}$  implies that for each $\alpha \in \Irrep(G^{R}),\tilde{T}(\alpha) \leq C	$ and $w	\in W,$ we have that 
		\begin{equation*}
			\E_{e \in E_w}  \left[ \mathcal{C}(\tilde{T}(\alpha), \pi_e) \right] \leq  \frac{|\tilde{T}(\alpha)|^{2}t}{r}\leq \frac{\delta^{2}}{8|G|^{6+C}}.
		\end{equation*}
		Then, using essentially the same argument as in Lemma~\ref{soundness_almost_perfect_lemma} from the previous section we have that
		\begin{equation*} 
			\frac{\delta^{2}}{8|G|^{6}}	\leq  \E_{e \in E} \left[|G|^{C} \sum_{\substack{\alpha \in \Irrep(G^{R}), \mathcal{C}(\tilde{T}(\alpha),\pi_e)=0 \\ \beta \in \Irrep(G^{L}) \\ \dim(\alpha), \dim(\beta)\geq 2 \\ |\tilde{T}(\alpha)| \leq C\\ \tilde{T}_{\pi_e}^{uq}(\alpha)\subseteq \tilde{T}(\beta) \subseteq \tilde{T}_{\pi_e}(\alpha) }} \dim(\alpha) \|\widehat{g_F}(\alpha)\|_{HS}^{2} \dim(\beta)				 \|\widehat{h_F}(\beta)\|_{HS}^{2} \right],
		\end{equation*}
		or, equivalently 
		\begin{equation*} 
			\frac{\delta^{2}}{8|G|^{6+C}}	\leq   \E_{e	\in E} \left[\sum_{\substack{\alpha \in \Irrep(G^{R}), \mathcal{C}(\tilde{T}(\alpha),\pi_e)=0 \\ \beta \in \Irrep(G^{L}) \\ \dim(\alpha), \dim(\beta)\geq 2 \\ |\tilde{T}(\alpha)| \leq C\\ \tilde{T}_{\pi_e}^{uq}(\alpha)\subseteq \tilde{T}(\beta) \subseteq \tilde{T}_{\pi_e}(\alpha) }} \dim(\alpha) \|\widehat{g_F}(\alpha)\|_{HS}^{2} \dim(\beta)				 \|\widehat{h_F}(\beta)\|_{HS}^{2}\right].
		\end{equation*}
		But then by using Markov's inequality there is  $E' \subseteq E, |E'| \geq \frac{\delta^{2}}{16|G|^{6+C}}|E|$  such that for every $e \in E'$ we have that 
		\begin{equation}\label{final_bound_perfect_completness}
			\frac{\delta^{2}}{16|G|^{6+C}} \leq \sum_{\substack{\alpha \in \Irrep(G^{R}), \mathcal{C}(\tilde{T}(\alpha),\pi_e)=0 \\ \beta \in \Irrep(G^{L}) \\ \dim(\alpha), \dim(\beta)\geq 2 \\ |\tilde{T}(\alpha)| \leq C\\ \tilde{T}_{\pi_e}^{uq}(\alpha)\subseteq \tilde{T}(\beta) \subseteq \tilde{T}_{\pi_e}(\alpha) }} \dim(\alpha) \|\widehat{g_F}(\alpha)\|_{HS}^{2} \dim(\beta)				 \|\widehat{h_F}(\beta)\|_{HS}^{2}.
		\end{equation}
		Consider now the following randomized labelling of the starting UFG Label Cover instance.	For a vertex $w \in W$  we pick a label $\sigma_w$ as follows. First, we pick a representation $\alpha$ with probability $\dim(\alpha)\|\widehat{g_F}(\alpha)\|_{HS}^{2}$. Then, if we consider the decomposition $\alpha=(\alpha_1,\hdots,\alpha_R)$, we select $\sigma_w$ uniformly at random from the set
	
		\begin{equation*}
			\left \{d \in [R] \mid \dim(\alpha_d)\geq 2 \textrm{ and for every } i \in [r],  \mathscr{p}_i(d)=b_{I_w(i)}\right\}.
		\end{equation*}
		Observe that whenever $\widehat{g_F}(\alpha)\neq 0$ we can always make such a choice by Lemma~\ref{non_zero_coefficient_perfect_completness_lemma}. For a vertex $v \in V$  we pick a label $\sigma_v$ in an analogous way, i.e.,~we first pick a representation $\beta$ with probability $\dim(\beta)\|\widehat{h_F}(\beta)\|_{HS}^{2}$, and then $\sigma_v$ uniformly at random from the set $\tilde{T}(\beta)$.  Then, by the inequality (\ref{final_bound_perfect_completness}) for each $e \in E'$ we have that $\alpha,\beta,$ satisfy with probability at least $\frac{\delta^{2}}{16 |G|^{6+C}}$ the following properties
		
	\begin{itemize}
		\item $|\tilde{T}(\alpha)| \leq C$,
		\item $\mathcal{C}(\tilde{T}(\alpha),\pi_e)=0$,
		\item $\tilde{T}_{\pi_e}^{uq}(\alpha) \subseteq  \tilde{T}(\beta) \subseteq \tilde{T}_{\pi_e}(\alpha)$.
	\end{itemize}
	Let us now condition on this event happening and calculate the probability that the agreement test is passed along the edge $e$. Observe that since $ \tilde{T}(\beta) \subseteq \tilde{T}_{\pi_e}(\alpha)$ we have that $|\tilde{T}(\beta)| \leq |\tilde{T}(\alpha)| \leq C$. Furthermore, since $\mathcal{C}(\tilde{T}(\alpha),\pi_e)=0$ we have that $\pi_e(\sigma_w) \in \tilde{T}_{\pi_e}^{uq}(\alpha)$ and hence since $\tilde{T}(\beta) \supseteq \tilde{T}_{\pi_e }^{uq}(\alpha)$ the probability that the test is passed is at least 
	\begin{equation*}
		\frac{1}{|\tilde{T}(\beta)|}	\geq \frac{1}{|\tilde{T}(\alpha)|}  \geq C^{-1}.
	\end{equation*}
	Hence, for $e \in E'$ 	the test passes with probability at least 
	\begin{equation*}
		\frac{\delta^{2}}{16 |G|^{6+C}} C^{-1},
	\end{equation*}
	and since $|E'| \geq \frac{\delta^{2}}{16 |G|^{6+C}} |E|$ the randomized assignment in expectation satisfies at least 
	\begin{equation*}
		\frac{\delta^{4}}{256 |G|^{12+2C}} C^{-1}
	\end{equation*}
	fraction of edges. But then observe that by the method of conditional expectations there is a labelling that satisfies at least $\frac{\delta^{4}}{256 |G|^{12+2C}}C^{-1}$ fraction of edges of the starting instance, which is a contradiction with the assumption that $\Opt(\Lambda_{UFG}) \leq \frac{\delta^{4}}{257 |G|^{12+2C}}C^{-1}$. Therefore  $\left|\E_{e \in E} \Theta^{e}(\textrm{low})\right|<  \frac{\delta}{2|G|^{3}}$ and our proof is complete.
	\end{proof}
\end{theorem}
\section*{Acknowledgements}
The second author is thankful to Johan H\aa stad for helpful discussions.
\bibliography{bibl}{}

\newcommand{\etalchar}[1]{$^{#1}$}
\begin{thebibliography}{KMOW17}

\bibitem[ABH21]{DBLP:conf/soda/AustrinBH21}
Per Austrin, Jonah Brown{-}Cohen, and Johan H{\aa}stad.
\newblock Optimal inapproximability with universal factor graphs.
\newblock In {\em Proc.\ $51$st Annual {ACM}-{SIAM} Symp.\ on Discrete
  Algorithms (SODA)}, pages 434--453, 2021.

\bibitem[AGH17]{AustrinGH17}
Per Austrin, Venkatesan Guruswami, and Johan H{\aa}stad.
\newblock (2+$\varepsilon$)-sat is {NP}-hard.
\newblock {\em SIAM Journal on Computing}, 46(5):1554--1573, 2017.

\bibitem[ALM{\etalchar{+}}98]{DBLP:conf/focs/AroraLMSS92}
Sanjeev Arora, Carsten Lund, Rajeev Motwani, Madhu Sudan, and Mario Szegedy.
\newblock Proof verification and the hardness of approximation problems.
\newblock {\em J. ACM}, 45(3):501–555, May 1998.

\bibitem[AS98]{DBLP:conf/focs/AroraS92}
Sanjeev Arora and Shmuel Safra.
\newblock Probabilistic checking of proofs: A new characterization of {NP}.
\newblock {\em J. {ACM}}, 45(1):70–122, January 1998.

\bibitem[BK20]{DBLP:journals/eccc/BhangaleK20}
Amey Bhangale and Subhash Khot.
\newblock Optimal inapproximability of satisfiable k-{LIN} over non-abelian
  groups.
\newblock {\em CoRR}, abs/2009.02815, 2020.
\newblock (to appear in STOC 2021).

\bibitem[BKO19]{BulinKO19}
Jakub Bul{\'{\i}}n, Andrei~A. Krokhin, and Jakub Oprsal.
\newblock Algebraic approach to promise constraint satisfaction.
\newblock In {\em Proc.\ $51$st ACM Symp.\ on Theory of Computing (STOC)},
  pages 602--613, 2019.

\bibitem[Cha16]{DBLP:journals/jacm/Chan16}
Siu~On Chan.
\newblock Approximation resistance from pairwise-independent subgroups.
\newblock {\em J. {ACM}}, 63(3):27:1--27:32, 2016.

\bibitem[Din07]{DBLP:journals/jacm/Dinur07}
Irit Dinur.
\newblock The {PCP} theorem by gap amplification.
\newblock {\em J. {ACM}}, 54(3):12, 2007.

\bibitem[EHR04]{DBLP:journals/tcs/EngebretsenHR04}
Lars Engebretsen, Jonas Holmerin, and Alexander Russell.
\newblock Inapproximability results for equations over finite groups.
\newblock {\em Theor. Comput. Sci.}, 312(1):17--45, 2004.

\bibitem[Fei02]{Feige02}
Uriel Feige.
\newblock Relations between average case complexity and approximation
  complexity.
\newblock In {\em Proc.\ $34$th ACM Symp.\ on Theory of Computing (STOC)},
  pages 534--543, 2002.

\bibitem[FGL{\etalchar{+}}96]{FeigeGLSS96}
Uriel Feige, Shafi Goldwasser, Laszlo Lov\'{a}sz, Shmuel Safra, and Mario
  Szegedy.
\newblock Interactive proofs and the hardness of approximating cliques.
\newblock {\em J. ACM}, 43(2):268–292, March 1996.

\bibitem[FJ12]{DBLP:conf/icalp/FeigeJ12}
Uriel Feige and Shlomo Jozeph.
\newblock Universal factor graphs.
\newblock In {\em Proc.\ $39$th International Colloq.\ of Automata, Languages
  and Programming (ICALP), Part I}, volume 7391 of {\em LNCS}, pages 339--350.
  Springer, 2012.

\bibitem[H{\aa}s01]{DBLP:journals/jacm/Hastad01}
Johan H{\aa}stad.
\newblock Some optimal inapproximability results.
\newblock {\em J. {ACM}}, 48(4):798--859, 2001.

\bibitem[Joz14]{Jozeph14}
Shlomo Jozeph.
\newblock Universal factor graphs for every {NP}-hard boolean csp.
\newblock In {\em Approximation, Randomization, and Combinatorial Optimization
  Algorithms and Techniques (APPROX/RANDOM)}, volume~28 of {\em LIPIcs}, pages
  274--283, 2014.

\bibitem[KMOW17]{KothariMOW17}
Pravesh~K Kothari, Ryuhei Mori, Ryan O'Donnell, and David Witmer.
\newblock Sum of squares lower bounds for refuting any csp.
\newblock In {\em Proc.\ $49$th ACM Symp.\ on Theory of Computing (STOC)},
  pages 132--145, 2017.

\bibitem[Rao11]{DBLP:journals/siamcomp/Rao11}
Anup Rao.
\newblock Parallel repetition in projection games and a concentration bound.
\newblock {\em SIAM Journal on Computing}, 40(6):1871--1891, 2011.

\bibitem[Raz98]{DBLP:journals/siamcomp/Raz98}
Ran Raz.
\newblock A parallel repetition theorem.
\newblock {\em SIAM Journal on Computing}, 27(3):763--803, 1998.

\bibitem[Ser77]{Serre1977}
Jean-Pierre Serre.
\newblock {\em Linear representations of finite groups}.
\newblock Springer, 1977.

\bibitem[Ter99]{terras_1999}
Audrey Terras.
\newblock {\em Fourier analysis on finite groups and applications}.
\newblock Number~43. Cambridge University Press, 1999.

\end{thebibliography}
\bibliographystyle{alpha}
\appendix
\section{Proof of Lemma~\ref{relationship_between_fourier_coeff_perfect_completeness}}
In this section we recall and then prove Lemma~\ref{relationship_between_fourier_coeff_perfect_completeness}.

\begin{lemma}[Lemma~\ref{relationship_between_fourier_coeff_perfect_completeness}, restated]
	Let $\alpha \in \Irrep(G^{R})$, $\alpha=\otimes_{d=1}^{R} \alpha_d$, and let $\pi\colon [R] \to [L]$. Consider the representation $\alpha\circ \tilde{\pi} \cong \oplus_{s=1}^{m} \tau^{s}$ of $G^{L}$ where each $\tau^{s} \in \Irrep(G^{L})$. Then for each $s$ we have that $\tilde{T}_{\pi}^{uq}(\alpha) \subseteq \tilde{T}(\tau^{s}) \subseteq \tilde{T}_{\pi}(\alpha)$.
	\begin{proof}
		Let $\ell \in [R]$ be such that $\pi(\ell) \in \tilde{T}_{\pi}^{uq}(\alpha)$, consider any block $\tau^s$ in the block diagonal matrix $\alpha \circ \tilde{\pi} \cong \oplus_{s=1}^{u} \tau^{s}$, and let $1\leq i',j'\leq \dim(\tau^{s})$ be arbitrary. Let $1\leq i,j \leq \dim(\alpha)$ be such that $[\alpha \circ \tilde{\pi}]_{ij}$ corresponds to $[\tau^{s}]_{i'j'}$, i.e.,~$i,j$ are unique such that $Q(\alpha,i,j) = (\tau^{s},i',j')$. Since $\alpha=\otimes_{d=1}^{R} \alpha_d$ we have that $[\alpha]_{ij} = \prod_{d=1}^{R}[\alpha_{d}]_{d_id_j},$ where $1\leq d_i,d_j \leq \dim(\alpha_d)$. But then
		\begin{equation*}
			[\tau^{s}]_{i'j'}({\bf x}) = [\alpha\circ \tilde{\pi}]_{ij} ({\bf x}) = \prod_{d=1}^{R}[\alpha_{d}]_{d_id_j} (x_{\pi(d)}) = \left(\prod_{d \in [R] \setminus \left \{\ell\right\}}[\alpha_{d}]_{d_id_j} (x_{\pi(d)})\right) \cdot [\alpha_{\ell}]_{\ell_i\ell_j}(x_{\pi(\ell)}).
		\end{equation*}
		 Let $[R]\setminus\left \{\ell\right\} = J_1 \sqcup J_2$ where 
		\begin{equation*}
			\begin{split}
				J_1&=\left \{ d \in [R] \mid d \neq \ell \textrm{ and } \pi(d)=\pi(\ell) \right\},\\
				J_2&=\left \{ d \in [R] \mid \pi(d)\neq\pi(\ell) \right\}.
			\end{split}
		\end{equation*}
		We can then write 
		\begin{equation*}
			[\tau^{s}]_{i'j'}({\bf x})	=  \left(\prod_{d \in J_1}[\alpha_{d}]_{d_id_j} (x_{\pi(\ell)})\right) \left(\prod_{d \in J_2}[\alpha_{d}]_{d_id_j} (x_{\pi(d)})\right)\cdot [\alpha_{\ell}]_{\ell_i\ell_j}(x_{\pi(\ell)}).
		\end{equation*}
		Let us write $\tau^{s}=\otimes_{d=1}^{L} \tau^{s}_d$ and for the sake of contradiction let us assume that $\dim(\tau^s_{\pi(\ell)})=1$.  By the equality above we have that 
		\begin{equation*}
			[\tau^s_{\pi(\ell)}]_{1,1}= \left(\prod_{d \in J_1}[\alpha_{d}]_{d_id_j} (x_{\pi(d)})\right)\cdot [\alpha_{\ell}]_{\ell_i\ell_j}(x_{\pi(\ell)}).
		\end{equation*}
		Observe that since $\pi(\ell) \in \tilde{T}_{\pi}^{uq}(\alpha)$ for each $d \in J_1$ we have $\dim(\alpha_d) = 1$.  Then by Lemma~\ref{orthogonality} we have 
		\begin{equation*}
			\begin{split}
				1 & = \frac{1}{\dim(\tau^{s}_{\pi(\ell)})}=\langle [\tau^s_{\pi(\ell)}]_{1,1},[\tau^s_{\pi(\ell)}]_{1,1}  \rangle_{L^{2}(G)} = \langle [\tau^{s}_{\pi(\ell)}]_{1,1}, \left(\prod_{d \in J_1}[\alpha_{d}]_{d_id_j}\right) [\alpha_{\ell}]_{\ell_i\ell_j}\rangle_{L^{2}(G)}	\\
				 & =\langle [\tau^{s}_{\pi(\ell)}]_{1,1}\left(\prod_{d \in J_1}\overline{[\alpha_{d}]}_{d_jd_i}\right), [\alpha_{\ell}]_{\ell_i\ell_j}\rangle_{L^{2}(G)} =0,
			\end{split}
		\end{equation*}
		which is a contradiction to $0\neq 1$. The last equality in the calculation above holds by Lemma~\ref{orthogonality}, since $[\tau^{s}_{\pi(\ell)}]_{1,1}\left(\prod_{d \in J_1}\right)\overline{[\alpha_{d}]}_{d_jd_i}$ is a one-dimensional representation and hence it is non-isomorphic to $[\alpha_{\ell}]$ which is of dimension at least $2$. 
		\par
		We now prove $\tilde{T}(\tau^{s}) \subseteq \tilde{T}_{\pi}(\alpha)$. Let us fix $k \in[L] \setminus \tilde{T}_{\pi}(\alpha)$, take any $s=1,\hdots,m,$ and let $1\leq i',j'\leq \dim(\tau^{s})$ be arbitrary. Furthermore, let us define $i,j$ as before. We have that 
		\begin{equation*}
			[\tau^{s}]_{i'j'}({\bf x})	=  \left(\prod_{\substack{d \in [R]\\ \pi(d) \neq k }}[\alpha_{d}]_{d_id_j} (x_{\pi(d)})\right)\left(\prod_{\substack{d \in [R]\\ \pi(d) = k }}[\alpha_{d}]_{d_id_j} (x_{\pi(d)})\right).
		\end{equation*}
		But then 
		\begin{equation*}
			\tau^{s}_k(x_{k}) = \prod_{\substack{d \in [R]\\ \pi(d) = k }}[\alpha_{d}]_{d_id_j},
		\end{equation*}
		 and hence $\tau^{s}_k$ is a one-dimensional representation as a product of one-dimensional representations. This shows that if $k \not \in \tilde{T}_{\pi}(\alpha)$  then $k \not \in \tilde{T}(\tau^{s})$, which is equivalent to  $\tilde{T}(\tau^{s}) \subseteq \tilde{T}_{\pi}(\alpha)$, which is what we wanted to prove.
	\end{proof}
\end{lemma}

\section{Proof of Lemma~\ref{smoothness_bhangale_khot}}
In this section we recall and then prove Lemma~\ref{smoothness_bhangale_khot}.
\begin{lemma}[Lemma~\ref{smoothness_bhangale_khot}, restated]
	Consider $(r,t)$-smooth UFG Label Cover $\Lambda_{UFG}$ constructed by the parallel repetition of $(1,0.51)$-UFG-NP-hard Max-TSA instance. There is a constant $d_0 \in (0,1/3)$ such that given any fixed $t$ for all sufficiently big $r$  we have that
	\begin{equation*}
		\Pr_{e \in E_w} [ |\pi_e(S)| <|S|^{d_0} ] \leq  \frac{1}{|S|^{d_0}}, \quad (\forall w \in W, S \subseteq  [R] ).
	\end{equation*}
	\begin{proof}
		We prove the statement for $d_0 =1/4$. Since every TSA constraint has $2^{5}=32$ possible assignments to the variables, and the verifier sends one binary variable from $t$ constraints to the second prover, the map $\pi_e\colon [R] \to [L]$ is $16^{t}$-to-$1$ for any $e \in E$. Hence, for each $S \subseteq [R]$ and every $e \in E$ we have that $|\pi_e(S)| \geq \frac{|S|}{16^{t}}$. But then if $|S| \geq 16^{\frac{4}{3} t}$ for any $w \in W$ we have that 
		\begin{equation*}
			\begin{split}
			\Pr_{e \in E_w} [ |\pi_e(S)| <|S|^{\frac{1}{4}} ] \leq \Pr_{e \in E_w} \left[ \frac{|S|}{16^{t}} <|S|^{\frac{1}{4}} \right] = \Pr_{e \in E_w} \left[ \frac{|S|^{\frac{3}{4}}}{16^{t}} <1 \right] \leq \Pr_{e \in E_w} \left[ \frac{(16^{\frac{4}{3}t } )^{\frac{3}{4}}}{16^{t}} <1 \right] \\ = \Pr_{e \in E_w} [ 1 <1 ] =0.
			\end{split}
		\end{equation*}
		It remains to consider the case when $|S|<16^{\frac{4}{3} t}$. In this case for any $w \in W$ we have
		\begin{equation*}
			\Pr_{e \in E_w} [ |\pi_e(S)| <|S|^{1/4} ] \leq 			\Pr_{e \in E_w} [ |\pi_e(S)| <|S| ] =  \E[\mathcal{C}(S,\pi_e)] \leq \frac{|S|^{2}t}{r} \leq \frac{16^{\frac{8}{3}t}t}{r},
		\end{equation*}
		where in the second last inequality we used Lemma~\ref{smoothness_lemma}. But then for sufficiently big $r$  we have that 	
		\begin{equation*}
			\frac{16^{\frac{8}{3}t}t}{r} \leq 16^{-\frac{1}{3} t} \leq |S|^{-1/4},
		\end{equation*}
		which concludes the proof of the lemma.	
	\end{proof}
\end{lemma}

\end{document}